\documentclass[10pt,aps,prl,twocolumn,showpacs,superscriptaddress]{revtex4-2} 

\usepackage[utf8]{inputenc}
\usepackage{graphicx}
\usepackage{amsmath}
\usepackage{amssymb}
\usepackage{bm}
\usepackage{txfonts}
\usepackage{multirow}
\usepackage{mathptmx} 
\usepackage{siunitx}
\usepackage{xargs}                    
\usepackage[pdftex,dvipsnames]{xcolor}
\usepackage[colorlinks=true, urlcolor=blue, linkcolor=blue, citecolor=blue]{hyperref}
\usepackage{xcolor}
\usepackage{braket}
\usepackage{soul}
\usepackage{verbatim}

\begin{document}
\title{Breaking the entangling gate speed limit for trapped-ion qubits using a phase-stable standing wave}
\date{\today}
\author{S. Saner$^{*}$, O. B\u{a}z\u{a}van$^{*}$, M. Minder, P. Drmota, D. J. Webb,  G. Araneda, R. Srinivas, D. M. Lucas, C. J. Ballance\\
\normalsize{Department of Physics, University of Oxford, Clarendon Laboratory, Parks Road, Oxford OX1 3PU, United Kingdom}\\
\normalsize{$^*$These authors contributed equally.\\
Email: sebastian.saner@physics.ox.ac.uk, oana.bazavan@physics.ox.ac.uk}}

\begin{abstract}

All laser-driven entangling operations for trapped-ion qubits have hitherto been performed without control of the optical phase of the light field, which precludes independent tuning of the carrier and motional coupling. 
By placing $^{88}$Sr$^+$ ions in a $\lambda=\SI{674}{\nano \meter}$ standing wave, whose relative position is controlled to $\approx\lambda/100$, we suppress the carrier coupling by a factor of 18, while coherently enhancing the spin-motion coupling.
We experimentally demonstrate that the off-resonant carrier coupling imposes a speed limit for conventional traveling-wave M\o{}lmer-S\o{}rensen gates; we use the standing wave to surpass this limit and achieve a gate duration of \SI{15}{\micro \second}, restricted by the available laser power.

\end{abstract}

\maketitle
Controlled light-matter interactions are essential for quantum computing \cite{cirac1995quantum, monroe1995demonstration, wineland1998experimental}, quantum simulation \cite{ blatt2012quantum, jaksch2005cold}, and metrology \cite{schmidt2005spectroscopy, wolf2016non}. 
For trapped ions, these applications typically require carrier interactions that only couple internal qubit states,
as well as sideband interactions that couple these internal states to their collective motion \cite{wineland1998experimental}. For example, the sideband interactions, driven by the spatial gradient of the carrier coupling, are used to mediate spin-spin interactions such as entangling gates \cite{blatt2008entangled}.
Conventionally, coherent control of laser-ion interactions is achieved using traveling waves (TWs) \cite{wineland1998experimental}. As the ions experience an averaged electric field and gradient over the interaction duration, the ratio between carrier coupling and sideband coupling is fixed. In contrast, the coupling strengths for ions in a standing wave (SW) vary with the spatial structure of the light field along its propagation direction. Consequently, the phase of the SW at the ions sets the ratio between the carrier and sideband coupling. Coherent SW interactions on a single ion have been studied previously using cavities \cite{mundt2002coupling, delaubenfels2015modulating}, integrated optics \cite{vasquez2022control} and free-space approaches \cite{schmiegelow2016phasestable}. However, coherent operations on multiple ions with a SW have so far been unexplored.

The tunability of the carrier:sideband coupling ratio is especially important for strong interactions where off-resonant terms start participating significantly and cannot be eliminated adiabatically. 
For example, in the conventional M\o{}lmer-S\o{}rensen (MS) mechanism \cite{sorensen2000entanglement}, the TW that generates the spin-motion coupling also gives rise to an off-resonant carrier coupling, which causes an error in the entangling operation.
This error becomes significant as the carrier interaction strength approaches the motional frequency, placing a limit on the speed of the entangling operation. Using a SW instead enables high-fidelity entangling operations that can surpass this speed limit by selectively enhancing the spin-motion coupling while coherently suppressing the detrimental carrier term \cite{mehta2019towards}. 
Fast entanglement generation is important for increasing the clock speed in trapped-ion quantum processors \cite{wineland1998experimental, Kielpinski2002architecture, schafer2018fast} and could enable experimental studies of vacuum entanglement and the propagation of quantum correlations in ion chains \cite{reznik2005violating, retzker2005detecting}. Furthermore, being able to tune the carrier:sideband ratio as a function of the position unlocks opportunities in metrology, such as sensing beyond the diffraction limit \cite{drechsler2021optical,qian2021super} or suppressing dipole light shifts when probing quadrupole clock transitions \cite{yudin2010hyper}. Standing waves may also be used for deterministic generation of entanglement in a quantum network~\cite{vetlugin2022deterministic}.

In this Letter, we use a free-space, phase-stabilized SW to implement single- and two-qubit gates. The SW is formed by two superimposed counter-propagating 674-nm beams that couple to the quadrupole qubit transition, \(5S_{1/2} \leftrightarrow 4D_{5/2}\), in \textsuperscript{88}Sr\textsuperscript{+}. 
The single-qubit gate is created using a monochromatic SW on resonance with the qubit transition while placing the node(s) of the SW at the position of the ion(s).
The two-qubit entangling gate is implemented via an MS-type scheme where we use a bichromatic SW instead of the conventional bichromatic TW. We show that the presence of the carrier term, in the context of the TW-MS gate, leads to a reduction in the spin-dependent force (SDF) magnitude, which scales with the Rabi frequency of this detrimental term, posing an inherent speed limit for this mechanism. Using the SW-MS instead, with the anti-nodes placed at the ions, we strongly suppress the undesired carrier term and show that we can surpass this speed limit.

To understand the interaction between a string of ions and a monochromatic SW driving a quadrupole transition, we consider two counter-propagating beams with Rabi frequency $\Omega$, detuning $\delta$ from the qubit resonance, and a tuneable phase difference $\Delta \phi = \phi_1 -\phi_2$ that is common to all (equally-spaced) ions in the chain \cite{cirac1993preparation,cirac1994quantum,wu1997jaynes}. The resulting interaction is
\begin{equation}\label{eq:monochrmatic_sw}
\begin{split}
    \hat{H}_\mathrm{SW} =\ & e^{-i\delta t}\hbar\eta\Omega\hat{S}_{+} e^{i\tilde{\phi}} (\hat{a}e^{-i \omega_z t} + \hat{a}^\dagger e^{i \omega_z t}) \cos\left(\Delta\phi/2\right) \\
    &+ e^{ -i\delta t}\hbar\Omega\hat{S}_{+} e^{i\tilde{\phi}} \sin\left(\Delta\phi/2\right) + \mathrm{h.c.},
\end{split}
\end{equation}
where $\eta$ denotes the Lamb-Dicke factor, the average phase $\tilde{\phi} = (\phi_1+\phi_2 + \pi)/2$, the spin-operator \footnote{$\hat{\sigma}_{\alpha}^{(i)} =\underbrace{\hat{\mathbb{I}}_2 \otimes ...\otimes \hat{\mathbb{I}}_2}_{i-1} \otimes \hat{\sigma}_{\alpha}\otimes \underbrace{\hat{\mathbb{I}}_2 \otimes ...\otimes \hat{\mathbb{I}}_2}_{n-i}$, where $\alpha \in \{+, -, x, y\}$ in this work} for $n$ ions is $\hat{S}_{+} = \sum_{i=1}^{n}\hat{\sigma}_{+}^{(i)} $
 and $\hat{a}^\dagger$~($\hat{a}$) denotes the creation (annihilation) operator of the motional mode~\cite{supplementary}. This expression is in the interaction picture w.r.t. the qubit frequency $\omega_0$, and the motional mode frequency $\omega_z$, after the rotating wave approximation
w.r.t. $\omega_0$.
By setting $\delta = 0$ or $\delta = \pm\omega_z$, we can bring the carrier or sidebands into resonance, respectively.
With $\Delta\phi$ the SW has an additional degree of freedom compared to the TW: by setting $\Delta\phi= 0$ we can drive the first sidebands while suppressing all even terms in the Lamb-Dicke expansion \cite{supplementary}, including the carrier term.
Conversely, if we set $\Delta\phi=\pi$ we drive the carrier coupling and suppress all odd terms in the Lamb-Dicke expansion, including the first sidebands. 

The MS interaction requires two tones symmetrically detuned about the qubit resonance by $\delta \approx \pm \omega_z$. To construct the Hamiltonian for a SW-MS interaction, we combine two monochromatic SWs as described by Eq.~\eqref{eq:monochrmatic_sw}, resulting in the bichromatic SW interaction
\begin{equation}
    \begin{split}
    \hat{H}_\mathrm{SW-MS} =\ & 2\hbar\eta\Omega \hat{S}_{\tilde{\phi}} \cos{(\delta t)}(\hat{a}e^{-i \omega_z t} + \hat{a}^\dagger e^{i \omega_z t})\cos{\left(\Delta\phi/2 \right)}\\    
     &+2\hbar \Omega\hat{S}_{\tilde{\phi}} \cos{(\delta t)} \sin{ \left(\Delta\phi/2 \right),}
    \end{split}
\label{eq:H_MS_sw}
\end{equation}
\noindent 
where the spin-operator for $n$ ions is $\hat{S}_{\tilde{\phi}} = \sum_{i=1}^{n}\hat{\sigma}_{\tilde{\phi}}^{(i)} $ with $\hat{\sigma}_{\tilde{\phi}}^{(i)} = \hat{\sigma}_x^{(i)}\cos\tilde{\phi}  +  \hat{\sigma}_y^{(i)}\sin\tilde{\phi}$ and the phase $\tilde{\phi} = (\tilde{\phi}_{\mathrm{BD}} + \tilde{\phi}_{\mathrm{RD}})/2$ is the mean optical phase between the blue- (BD) and the red- (RD) detuned SWs. Further, we assume that the BD and RD SWs are in phase at the position of the ion(s), i.e. $\Delta\phi_\textrm{BD} = \Delta\phi_\textrm{RD}= \Delta\phi$. The first term corresponds to a spin-dependent force (SDF) and the second term drives the carrier transition off-resonantly. Notably, these terms commute.
Similar to the monochromatic SW,  we can drive the motional coupling while suppressing the spurious carrier coupling by setting $\Delta\phi = 0$.

The advantage of using a SW-MS interaction becomes evident when considering the conventional MS scheme, which consists of a BD and RD TW at $\delta \approx \pm \omega_z$:
\begin{equation}
    \begin{split}
    \hat{H}_\mathrm{TW-MS} =\ & \hbar\eta\Omega \hat{S}_{\phi} \cos{(\delta t)} (\hat{a}e^{-i \omega_z t} + \hat{a}^\dagger e^{i \omega_z t})\\
    &+\hbar \Omega \hat{S}_{\phi-\pi/2} \cos{(\delta t)},
    \end{split}
\label{eq:H_MS}
\end{equation}
where $\phi$ is the mean optical phase between the BD and RD TWs.
Crucially, in this case, the carrier and the SDF terms no longer commute.
Hence, when using this SDF to implement a two-qubit entangling gate, the off-resonant carrier coupling introduces an error, which increases with $\Omega$. This error can be partially mitigated by adiabatic ramping of the interaction (i.e., amplitude pulse shaping), which ensures a smooth transition into the interaction picture w.r.t. the carrier coupling if $\Omega \lesssim \delta$. Nevertheless, the non-commuting carrier term effectively limits the speed of entangling operations because it saturates the achievable SDF magnitude. By moving into the interaction picture w.r.t. the carrier term \cite{ Roos2008ion,sutherland2019versatile, Bazavan2022synthesizing}, Eq.~\eqref{eq:H_MS} becomes
\begin{equation}\label{eqn:H_I}
    \begin{split}
            \hat{H}_\mathrm{TW-MS}^I &=  \hbar \Omega_{\rm SDF}  \cos{(\delta t)} \hat{S}_\phi(\hat{a} e^{-i \omega_z t} + \hat{a}^\dagger e^{i \omega_z t}),\\
            \Omega_{\rm SDF}(\Omega,\delta) &= \eta\Omega [J_0(2\Omega/\delta)+ J_2(2\Omega/\delta)],
    \end{split}
\end{equation}
where $J_0$ and $J_2$ are Bessel functions of the first kind. The effective coupling strength $\Omega_\textrm{SDF}$ has a global maximum which limits the gate speed even if $\Omega$ is further increased (e.g.~by increasing the laser power).
 
\begin{figure}
\centering\includegraphics{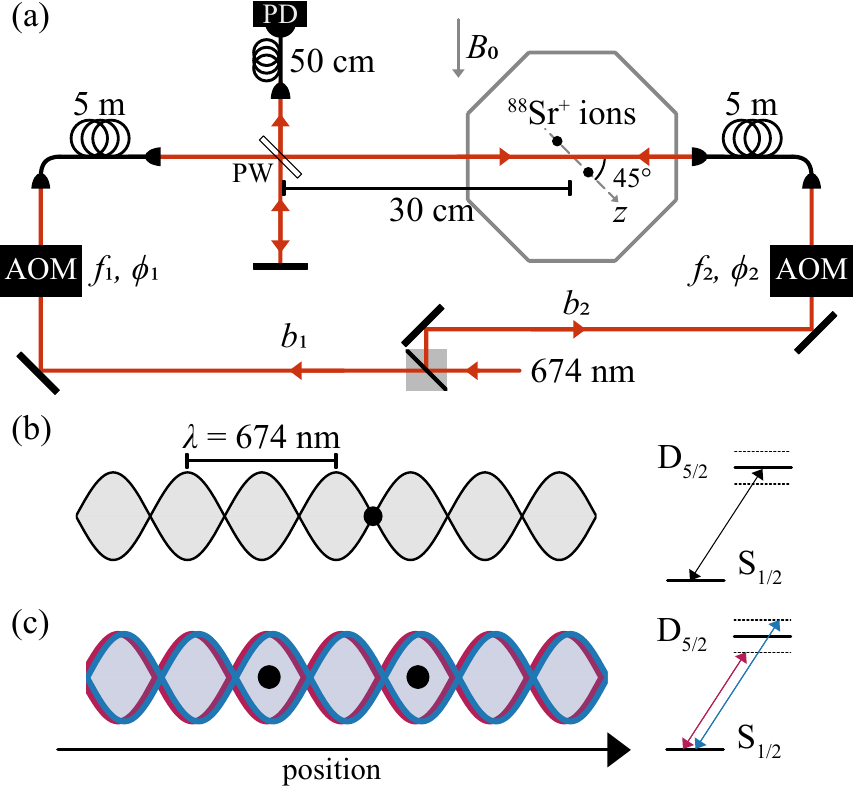}\\
\caption{
    (a) Schematic of the experimental apparatus. The incoming 674-nm beam is split into two beams ($b_1$, $b_2$). The acousto-optic modulators (AOMs) are used to control the frequencies ($f_1$, $f_2$) and phases ($\phi_1$, $\phi_2$) of the two counter-propagating beams, which have  polarization parallel to $B_0$ and equal intensities at the ions. 
    We close the resulting interferometer with a pick-off window (PW) $\approx \SI{30}{\centi \meter}$ away from the ion(s). 
    For fast feedback (see text) we stabilize the interference fringe intensity on a photodiode (PD) by adjusting $\phi_1$.
    (b) Monochromatic resonant SW for single-qubit rotations. (c) Bichromatic off-resonant SW for two-qubit gates.}
    \label{fig:setup}
\end{figure}

We experimentally compare single- and two-qubit operations implemented via SWs or TWs using the setup shown in Fig.~\ref{fig:setup}(a). We trap one or two $^{88}$Sr$^+$ ions in a 3D radio-frequency Paul trap~\cite{thirumalai2019high, schafer2018fastthesis} with a quantization axis defined by a magnetic field $B_0$. Our qubit is encoded in ${\ket{\downarrow}\equiv\ket{5S_{1/2},\,m_j = -\frac{1}{2}}}$ and $\ket{\uparrow}\equiv\ket{4D_{5/2},\,m_j = -\frac{3}{2}}$; we address the quadrupole qubit transition using a 674-nm laser. 
The laser output is split into two beams, $b_1$ and $b_2$.
Both beams have a $\approx\SI{21}{\micro\meter}$ waist radius at the ion position.
For experiments with a TW we use $b_1$ alone.
To generate a free-space SW, light from both beams is aligned in a counter-propagating geometry onto the ions. The beams make an angle of $\approx45^\circ$ to the trap $z$-axis resulting in an ion separation projected on the SW axis of $\approx\SI{3.8}{\micro \meter}\cdot\cos(45^\circ) = 4\lambda$.

To perform coherent operations with the SW, we need to control the phase $\Delta\phi$ at the position of the ion(s), which is achieved by adjusting phase $\phi_1$. 
We increase the passive stability of $\Delta\phi$ with an enclosure around the free-space optical paths. Additionally, 
we actively stabilize $\Delta\phi$ on two time scales: fast feedback derived from optical interference sampled near the position of the ions, and slow feedback derived from Ramsey experiments on the ion(s)~\cite{supplementary}.
Using a single ion as a sensor, we observe residual phase fluctuations with an rms deviation of $\approx\SI{0.12}{\radian}$ (position fluctuations of $\approx\lambda/100$) over one hour. This is near the shot noise limit, i.e. $\SI{0.10}{\radian}$ for 100 shots of feedback.

\begin{figure}
\includegraphics{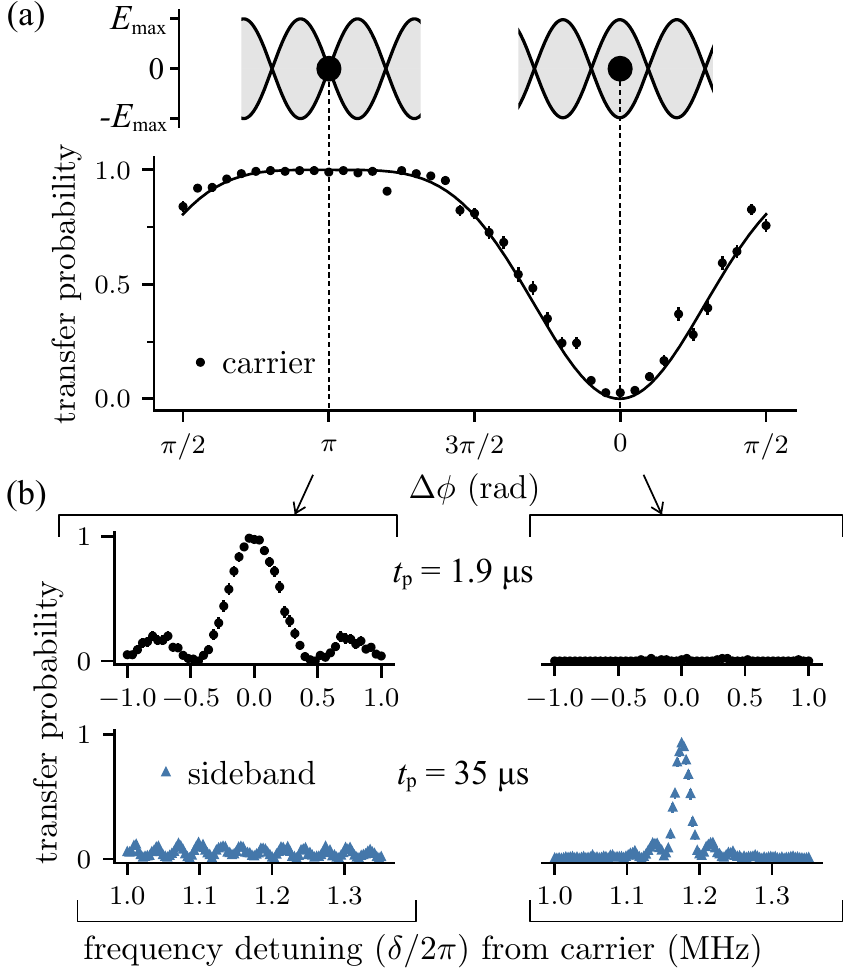}
\caption{Monochromatic SW interacting with a single ion. (a) Qubit state transfer probability as a function of the SW phase at the ion position, while the SW is resonant with the carrier.
We indicate the ion positions in the SW that maximize ($\Delta\phi = \pi$) or minimize ($\Delta\phi = 0$) the carrier coupling for a quadrupole transition.
The SW pulse duration $t_p$ is set such that complete population transfer is achieved at maximal carrier coupling. 
(b) Detuning scans over carrier (circles) and motional sideband (triangles) resonance while placing the ion at a field node (left column) or field anti-node (right column). For each resonance, $t_p$ is chosen such that full population transfer is reached in the case of maximal coupling to the SW.}
\label{fig:lattice-single-rotations}
\end{figure}

We probe the position of the SW relative to a single ion by applying a monochromatic SW pulse on resonance with the qubit transition [Figs.~\ref{fig:setup}(b),~\ref{fig:lattice-single-rotations}(a)].
The pulse duration corresponds to a $\pi$-pulse at maximum carrier coupling. As we are driving an electric quadrupole transition, this maximum occurs at the nodes of the SW, where the gradient of the electric field has the largest amplitude~\cite{mundt2002coupling}. Conversely, the sideband coupling is maximised at the anti-nodes of the SW as it is proportional to the spatial derivative of the carrier coupling along the motional direction. Hence, we can maximize the carrier and minimize the sideband coupling, or vice versa, by selecting $\Delta\phi = \pi$ or $\Delta\phi = 0$ [Fig.~\ref{fig:lattice-single-rotations}(b)]. The transfer probability shown in Fig.~\ref{fig:lattice-single-rotations}(a) has
a quartic dependence on $\Delta\phi$ near $\Delta\phi = \pi$ and a quadratic dependence
near $\Delta\phi = 0$~\cite{supplementary}.
When probing the suppressed motional sideband [Fig.~\ref{fig:lattice-single-rotations}(b) left], we observe only features that are due to the off-resonant (by $\approx \SI{1.2}{\mega \hertz}$) carrier coupling.
By changing $\Delta\phi$, we can realize any ratio between carrier and sideband coupling.

We measure Rabi frequencies by scanning the SW pulse duration at the carrier resonance, for both $\Delta\phi=\pi$ and $\Delta\phi=0$. We observe this ratio to be $\num{18}$, corresponding to a suppression of $\SI{25}{\dB}$ between maximal and minimal carrier coupling.
This suppression is consistent with the measured interferometric stability and the residual power imbalance between $b_1$ and $b_2$.
Furthermore, we perform randomized benchmarking \cite{knill2008randomized} to evaluate the quality of single-qubit gates implemented using the SW and TW with the same duty cycle. We obtain errors of $\num{1.44(3) e-3}$ and $\num{1.73(3) e-3}$ per Clifford gate, respectively. Thus, use of the SW is not detrimental to single-qubit control.

Next, we experimentally investigate the saturation effect caused by the non-commuting carrier coupling [Eq.~\eqref{eq:H_MS}] when generating an SDF with a bichromatic TW, and compare it to the SDF generated by a bichromatic SW. To create the TW bichromatic field, we apply two tones to the AOM in $b_1$, while for the SW we apply the same two tones in both beams, $b_1$ and $b_2$. These tones are symmetrically detuned by $\delta\approx\pm\omega_z$ from the qubit resonance. This results in an SDF on the axial mode ($\omega_z/2\pi\approx \SI{1.2}{\mega\hertz}$) of a single ion. 
We extract its strength $\Omega_\textrm{SDF}(\Omega, \delta)$ by applying the SDF for variable durations \cite{Bazavan2022synthesizing}.
We used an adiabatic ramp duration of $\SI{3.6}{\micro \second}$ for these measurements~\cite{fnpulseshape}.

\begin{figure}
    \centering
    \includegraphics{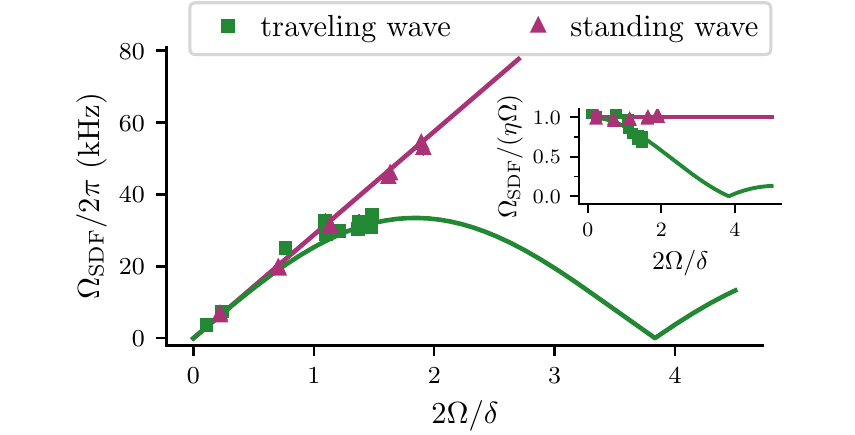}
    \caption{Spin-dependent force magnitude $\Omega_{\mathrm{SDF}}$ (normalized by $\eta\Omega$ in the inset) versus $2\Omega/\delta$, as measured for a single ion with $\eta= 0.051$. We extract $\Omega_{\mathrm{SDF}}$ by applying a conventional bichromatic TW field (squares), or a bichromatic SW field (triangles), for variable durations. The solid lines show the analytical dependence; as predicted by the theory and shown explicitly in the inset, the TW coupling follows the Bessel functions ($|J_0+J_2|$), while the SW coupling remains constant~\cite{fninterference}.
    }
    \label{fig:J0J2}
\end{figure}

For the TW, we observe a coupling that scales with the expected Bessel function dependence $|J_0(2\Omega/\delta) + J_2(2\Omega/\delta)|$ [Eq.~\eqref{eqn:H_I},~Fig.~\ref{fig:J0J2}]. Hence, when using the TW, there exists a maximum achievable interaction strength that imposes a speed limit on the interaction regardless of the available laser power. This limit is caused by the increasingly strong off-resonant non-commuting carrier excitation and not by technical aspects such as pulse shaping. 
For the SW, we demonstrate that no~such speed limit exists. We place the ion at the maximum intensity of both the RD and BD SWs~\cite{supplementary} and observe that the interaction magnitude~\cite{fninterference} 
increases linearly with $\Omega$~(Fig.~\ref{fig:J0J2}).

An important application of a bichromatic SW is to generate strong SDFs without any off-resonant carrier excitation. 
This can then be combined with pulse-segmentation techniques~\cite{Steane2014pulsed, palermo2017fast, schafer2018fast} to enable fast, non-adiabatic entangling operations.
Additionally, undesired squeezing terms $\mathcal{O}(\eta^{2})$, which were the dominant source of error in the fastest previous implementation \cite{schafer2018fast}, are suppressed~\cite{supplementary}.

\begin{figure}
    \centering
  \includegraphics{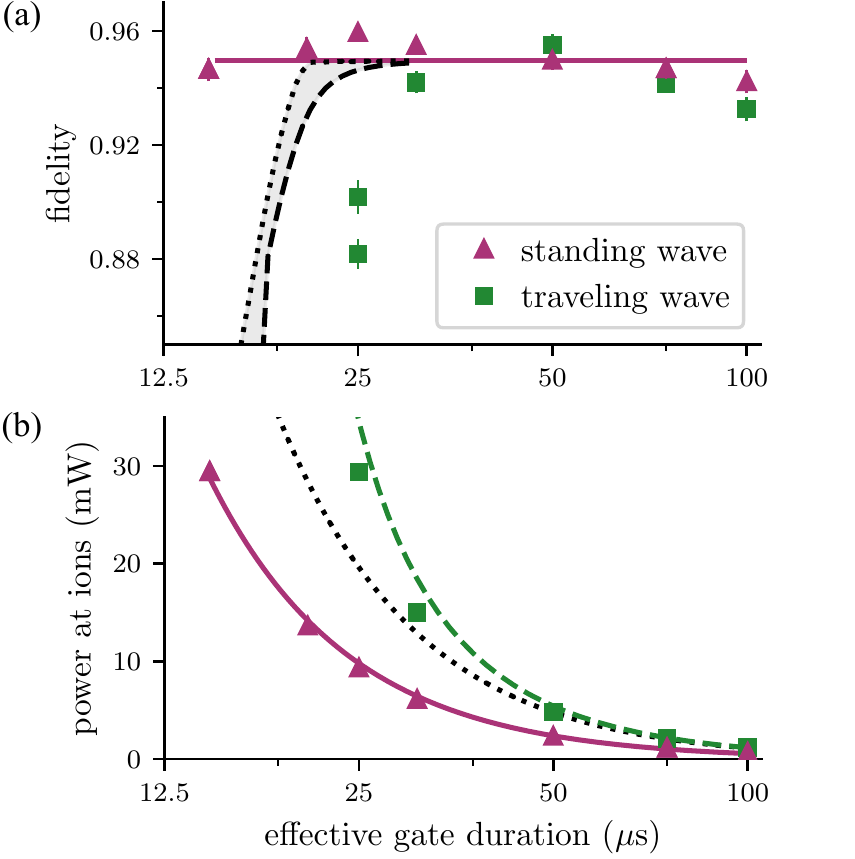}\\
    \caption{Characterization of SW (triangles) and TW (squares) M\o{}lmer-S\o{}rensen gates as a function of the effective two-qubit gate duration $(2\pi/\delta_g)$.
    (a) Using the SW, we achieve gate fidelities that are consistent with $\approx 0.95$ (solid line) for all gate durations. Using the TW the fidelity decreases rapidly for durations $\leq \SI{25}{\micro\second}$. As a guide to the eye, we show TW-MS simulations (dotted and dashed lines), with the maximum fidelity normalized to $0.95$.
    (b) Total laser power required at the ions to generate the gate interaction. The constructive interference of the SW reduces the required power at the ions by a factor two. The solid curve shows an inverse-square fit to the SW data. We scale this curve by a factor of two (dotted line) for comparison with the TW data.
    At fast gate durations $\left(\lesssim\SI{40}{\micro\second}\right)$, the required power for TW gates exceeds this prediction as a result of the saturation of the SDF (Fig.~\ref{fig:J0J2}). We scale this prediction by the expected Bessel function dependence (dashed line) and find good agreement with the measurements.
    }
    \label{fig:standing_wave_MS}
\end{figure}

We experimentally demonstrate two-qubit MS gates using a bichromatic TW for gate speeds in a regime where the carrier coupling induces a significant error which cannot be eliminated adiabatically. However, the bichromatic SW enables us to surpass this limit without degradation of the fidelity (Fig.~\ref{fig:standing_wave_MS}). To implement the SW-MS gate, we simultaneously suppress the carrier coupling on both ions by adjusting the ion spacing such that they are both located at anti-nodes of the SW [Fig.~\ref{fig:setup}(c)] \cite{supplementary}. 
We perform the TW and SW two-qubit entangling gates on the axial in-phase mode and optimize the experimental parameters to maximize the Bell-state fidelity for a fixed gate duration. In both cases, we use a ramp duration of $\SI{10}{\micro\second}$ to minimize coupling to the other motional modes~\cite{fnpulseshape}.
This pulse ramping could be replaced with more sophisticated amplitude shaping techniques \cite{Steane2014pulsed, schafer2018fast, palermo2017fast}. 

In Fig.~\ref{fig:standing_wave_MS}(a) we show the two-qubit fidelities achieved with the two schemes as a function of the effective gate duration ($2\pi/\delta_g$, where $\delta_g= \delta - \omega_z$) \footnote{The effective gate duration is approximately equal to the full-width half maximum (FWHM) of the pulse shape. The pulse duration from start to end is $2\pi/\delta_g + t_R$, where $t_R$ is the ramp duration. The ramp shape is as defined in \cite{fnpulseshape}.}. For slower gates, the fidelity of the SW-MS is comparable with that of the TW-MS. For faster gates, the fidelity of the TW-MS degrades rapidly. This is also predicted by direct numerical integration of Eq.~\eqref{eq:H_MS}; we set all the parameters to the experimental values except for the Rabi frequency $\Omega$, which we optimize for maximum fidelity (dashed line). We also indicate the idealized case which neglects imperfect transfer into the interaction picture w.r.t.~the carrier [Eq.~\eqref{eq:H_MS}] (dotted line). We believe that the measured fidelities degrade sooner (by $\approx \SI{5}{\micro\second}$) as a result of experimental imperfections (e.g.~in ramp shape) not captured in the numerical model. 
In contrast, the fidelity for the SW-MS is consistent with  $\approx 0.95$ over the entire available power range, showing that we have eliminated the limit arising from the carrier coupling. The shortest SW-MS gate was \SI{15}{\micro \second}, limited by the total available laser power of \SI{29}{\milli\watt}.

In Fig.~\ref{fig:standing_wave_MS}(b), we plot the total laser power delivered to the ions as a function of the effective gate duration. For the SW, the total required power ($b_1$ and $b_2$ summed) closely follows an inverse-square law. For a given duration, the TW-MS requires significantly more power than the SW-MS: the interference effect between the counter-propagating beams gives the SW-MS a factor 2 increase in power efficiency; the saturation effect (Fig.~\ref{fig:J0J2}) further increases the TW-MS power requirement.

We believe the main source of infidelity for entangling operations is phase noise from the 674-nm laser, which is common to both gate implementations. We estimate the sources of error that are introduced by the SW in the Supplemental Material~\cite{supplementary}: the visibility error due to amplitude imbalance between beams $b_1$ and $b_2$; the quality of the SW phase stabilization, which introduces a position jitter of the SW relative to the ions; mismatched spacing of the ions relative to the SW periodicity; and phase misalignment of the BD and RD SWs ($\Delta\phi_\mathrm{BD}\neq0\ {\rm or}\ \Delta\phi_\mathrm{RD}\neq0$).
The total error introduced is $<\num{9e-3}$ for a square pulse and $< \num{2e-5}$ when using a shaped pulse, which suppresses carrier related errors by three orders of magnitude and is employed for the results presented in Fig.~\ref{fig:standing_wave_MS}.

In conclusion, we implemented single- and two-qubit operations for trapped-ion qubits using a phase-stabilized SW. 
Two counter-propagating beams create the SW, whose relative phase $\Delta\phi$ at the ion position is stable to $\approx \lambda/100$. This enabled us to tune the ratio of the field intensity and gradient that the ions experience, which sets the relative strengths of the sideband and carrier interactions.
We use this new degree of control to suppress the unwanted off-resonant carrier coupling (by a factor of 18), while coherently enhancing the motional coupling during two-qubit gates.
We show theoretically and experimentally that the non-commuting carrier term imposes a limit on the speed of conventional TW-MS gates, which we circumvented by using the SW-MS interaction.
These optical phase control techniques could also be applied in the previous Raman-based scheme~\cite{schafer2018fast}, where they could mitigate squeezing terms, which were the leading error source; we note that for the SW-MS those terms are inherently suppressed.
Our work shows a clear path towards entangling gates with durations shorter than the motional period of the ions ($\lesssim \SI{1}{\micro \second}$)
at wavelengths that are amenable to large-scale chip integration using standard integrated optics~\cite{wang2011laser, niffenegger2020integrated, Mehta2020}  
and without the technical challenges of using high-power blue Raman beams~\cite{schafer2018fast}, pulsed lasers~\cite{garcia2003speed, wong2017demonstration} or Rydberg schemes~\cite{zhang2020submicrosecond}.
\section{Acknowledgements}
We would like to thank V. M. Sch{\"a}fer, A. C. Hughes and D. P. Nadlinger for thoughtful discussions.
This work was supported by the US Army Research Office (W911NF-20-1-0038) and the UK EPSRC Hub in Quantum Computing and Simulation (EP/T001062/1). CJB acknowledges support from a UKRI FL Fellowship. RS acknowledges funding from the EPSRC Fellowship EP/W028026/1 and Balliol College, Oxford. GA acknowledges support from Wolfson College, Oxford.

\section{Competing Interests}
GA consults for Nu Quantum Ltd. RS is partially employed by Oxford Ionics Ltd. CJB is a director of Oxford Ionics Ltd. All other authors declare no competing interests.

\bibliographystyle{apsrev4-2}
\bibliography{bibliography}

\begin{thebibliography}{44}%
\makeatletter
\providecommand \@ifxundefined [1]{%
 \@ifx{#1\undefined}
}%
\providecommand \@ifnum [1]{%
 \ifnum #1\expandafter \@firstoftwo
 \else \expandafter \@secondoftwo
 \fi
}%
\providecommand \@ifx [1]{%
 \ifx #1\expandafter \@firstoftwo
 \else \expandafter \@secondoftwo
 \fi
}%
\providecommand \natexlab [1]{#1}%
\providecommand \enquote  [1]{``#1''}%
\providecommand \bibnamefont  [1]{#1}%
\providecommand \bibfnamefont [1]{#1}%
\providecommand \citenamefont [1]{#1}%
\providecommand \href@noop [0]{\@secondoftwo}%
\providecommand \href [0]{\begingroup \@sanitize@url \@href}%
\providecommand \@href[1]{\@@startlink{#1}\@@href}%
\providecommand \@@href[1]{\endgroup#1\@@endlink}%
\providecommand \@sanitize@url [0]{\catcode `\\12\catcode `\$12\catcode
  `\&12\catcode `\#12\catcode `\^12\catcode `\_12\catcode `\%12\relax}%
\providecommand \@@startlink[1]{}%
\providecommand \@@endlink[0]{}%
\providecommand \url  [0]{\begingroup\@sanitize@url \@url }%
\providecommand \@url [1]{\endgroup\@href {#1}{\urlprefix }}%
\providecommand \urlprefix  [0]{URL }%
\providecommand \Eprint [0]{\href }%
\providecommand \doibase [0]{https://doi.org/}%
\providecommand \selectlanguage [0]{\@gobble}%
\providecommand \bibinfo  [0]{\@secondoftwo}%
\providecommand \bibfield  [0]{\@secondoftwo}%
\providecommand \translation [1]{[#1]}%
\providecommand \BibitemOpen [0]{}%
\providecommand \bibitemStop [0]{}%
\providecommand \bibitemNoStop [0]{.\EOS\space}%
\providecommand \EOS [0]{\spacefactor3000\relax}%
\providecommand \BibitemShut  [1]{\csname bibitem#1\endcsname}%
\let\auto@bib@innerbib\@empty
\bibitem [{\citenamefont {Cirac}\ and\ \citenamefont
  {Zoller}(1995)}]{cirac1995quantum}%
  \BibitemOpen
  \bibfield  {author} {\bibinfo {author} {\bibfnamefont {J.~I.}\ \bibnamefont
  {Cirac}}\ and\ \bibinfo {author} {\bibfnamefont {P.}~\bibnamefont {Zoller}},\
  }\href {https://doi.org/https://doi.org/10.1103/PhysRevLett.74.4091}
  {\bibfield  {journal} {\bibinfo  {journal} {Phys. Rev. Lett.}\ }\textbf
  {\bibinfo {volume} {74}},\ \bibinfo {pages} {4091} (\bibinfo {year}
  {1995})}\BibitemShut {NoStop}%
\bibitem [{\citenamefont {Monroe}\ \emph {et~al.}(1995)\citenamefont {Monroe},
  \citenamefont {Meekhof}, \citenamefont {King}, \citenamefont {Itano},\ and\
  \citenamefont {Wineland}}]{monroe1995demonstration}%
  \BibitemOpen
  \bibfield  {author} {\bibinfo {author} {\bibfnamefont {C.}~\bibnamefont
  {Monroe}}, \bibinfo {author} {\bibfnamefont {D.~M.}\ \bibnamefont {Meekhof}},
  \bibinfo {author} {\bibfnamefont {B.~E.}\ \bibnamefont {King}}, \bibinfo
  {author} {\bibfnamefont {W.~M.}\ \bibnamefont {Itano}},\ and\ \bibinfo
  {author} {\bibfnamefont {D.~J.}\ \bibnamefont {Wineland}},\ }\href
  {https://doi.org/10.1103/PhysRevLett.75.4714} {\bibfield  {journal} {\bibinfo
   {journal} {Phys. Rev. Lett.}\ }\textbf {\bibinfo {volume} {75}},\ \bibinfo
  {pages} {4714} (\bibinfo {year} {1995})}\BibitemShut {NoStop}%
\bibitem [{\citenamefont {Wineland}\ \emph {et~al.}(1998)\citenamefont
  {Wineland}, \citenamefont {Monroe}, \citenamefont {Itano}, \citenamefont
  {Leibfried}, \citenamefont {King},\ and\ \citenamefont
  {Meekhof}}]{wineland1998experimental}%
  \BibitemOpen
  \bibfield  {author} {\bibinfo {author} {\bibfnamefont {D.~J.}\ \bibnamefont
  {Wineland}}, \bibinfo {author} {\bibfnamefont {C.}~\bibnamefont {Monroe}},
  \bibinfo {author} {\bibfnamefont {W.~M.}\ \bibnamefont {Itano}}, \bibinfo
  {author} {\bibfnamefont {D.}~\bibnamefont {Leibfried}}, \bibinfo {author}
  {\bibfnamefont {B.~E.}\ \bibnamefont {King}},\ and\ \bibinfo {author}
  {\bibfnamefont {D.~M.}\ \bibnamefont {Meekhof}},\ }\href
  {https://doi.org/10.6028/jres.103.019} {\bibfield  {journal} {\bibinfo
  {journal} {J. Res. Natl. Inst. Stand. Technol.}\ }\textbf {\bibinfo {volume}
  {103}},\ \bibinfo {pages} {259} (\bibinfo {year} {1998})}\BibitemShut
  {NoStop}%
\bibitem [{\citenamefont {Blatt}\ and\ \citenamefont
  {Roos}(2012)}]{blatt2012quantum}%
  \BibitemOpen
  \bibfield  {author} {\bibinfo {author} {\bibfnamefont {R.}~\bibnamefont
  {Blatt}}\ and\ \bibinfo {author} {\bibfnamefont {C.~F.}\ \bibnamefont
  {Roos}},\ }\href {https://doi.org/10.1038/nphys2252} {\bibfield  {journal}
  {\bibinfo  {journal} {Nat. Phys.}\ }\textbf {\bibinfo {volume} {8}},\
  \bibinfo {pages} {277} (\bibinfo {year} {2012})}\BibitemShut {NoStop}%
\bibitem [{\citenamefont {Jaksch}\ and\ \citenamefont
  {Zoller}(2005)}]{jaksch2005cold}%
  \BibitemOpen
  \bibfield  {author} {\bibinfo {author} {\bibfnamefont {D.}~\bibnamefont
  {Jaksch}}\ and\ \bibinfo {author} {\bibfnamefont {P.}~\bibnamefont
  {Zoller}},\ }\href
  {https://doi.org/https://doi.org/10.1016/j.aop.2004.09.010} {\bibfield
  {journal} {\bibinfo  {journal} {Ann. Phys.}\ }\textbf {\bibinfo {volume}
  {315}},\ \bibinfo {pages} {52} (\bibinfo {year} {2005})},\ \bibinfo {note}
  {special Issue}\BibitemShut {NoStop}%
\bibitem [{\citenamefont {Schmidt}\ \emph {et~al.}(2005)\citenamefont
  {Schmidt}, \citenamefont {Rosenband}, \citenamefont {Langer}, \citenamefont
  {Itano}, \citenamefont {Bergquist},\ and\ \citenamefont
  {Wineland}}]{schmidt2005spectroscopy}%
  \BibitemOpen
  \bibfield  {author} {\bibinfo {author} {\bibfnamefont {P.~O.}\ \bibnamefont
  {Schmidt}}, \bibinfo {author} {\bibfnamefont {T.}~\bibnamefont {Rosenband}},
  \bibinfo {author} {\bibfnamefont {C.}~\bibnamefont {Langer}}, \bibinfo
  {author} {\bibfnamefont {W.~M.}\ \bibnamefont {Itano}}, \bibinfo {author}
  {\bibfnamefont {J.~C.}\ \bibnamefont {Bergquist}},\ and\ \bibinfo {author}
  {\bibfnamefont {D.~J.}\ \bibnamefont {Wineland}},\ }\href
  {https://doi.org/10.1126/science.1114375} {\bibfield  {journal} {\bibinfo
  {journal} {Science}\ }\textbf {\bibinfo {volume} {309}},\ \bibinfo {pages}
  {749} (\bibinfo {year} {2005})}\BibitemShut {NoStop}%
\bibitem [{\citenamefont {Wolf}\ \emph {et~al.}(2016)\citenamefont {Wolf},
  \citenamefont {Wan}, \citenamefont {Heip}, \citenamefont {Gebert},
  \citenamefont {Shi},\ and\ \citenamefont {Schmidt}}]{wolf2016non}%
  \BibitemOpen
  \bibfield  {author} {\bibinfo {author} {\bibfnamefont {F.}~\bibnamefont
  {Wolf}}, \bibinfo {author} {\bibfnamefont {Y.}~\bibnamefont {Wan}}, \bibinfo
  {author} {\bibfnamefont {J.~C.}\ \bibnamefont {Heip}}, \bibinfo {author}
  {\bibfnamefont {F.}~\bibnamefont {Gebert}}, \bibinfo {author} {\bibfnamefont
  {C.}~\bibnamefont {Shi}},\ and\ \bibinfo {author} {\bibfnamefont {P.~O.}\
  \bibnamefont {Schmidt}},\ }\href
  {https://doi.org/https://doi.org/10.1038/nature16513} {\bibfield  {journal}
  {\bibinfo  {journal} {Nature}\ }\textbf {\bibinfo {volume} {530}},\ \bibinfo
  {pages} {457} (\bibinfo {year} {2016})}\BibitemShut {NoStop}%
\bibitem [{\citenamefont {Blatt}\ and\ \citenamefont
  {Wineland}(2008)}]{blatt2008entangled}%
  \BibitemOpen
  \bibfield  {author} {\bibinfo {author} {\bibfnamefont {R.}~\bibnamefont
  {Blatt}}\ and\ \bibinfo {author} {\bibfnamefont {D.}~\bibnamefont
  {Wineland}},\ }\href {https://doi.org/10.1038/nature07125} {\bibfield
  {journal} {\bibinfo  {journal} {Nature}\ }\textbf {\bibinfo {volume} {453}},\
  \bibinfo {pages} {1008} (\bibinfo {year} {2008})}\BibitemShut {NoStop}%
\bibitem [{\citenamefont {Mundt}\ \emph {et~al.}(2002)\citenamefont {Mundt},
  \citenamefont {Kreuter}, \citenamefont {Becher}, \citenamefont {Leibfried},
  \citenamefont {Eschner}, \citenamefont {Schmidt-Kaler},\ and\ \citenamefont
  {Blatt}}]{mundt2002coupling}%
  \BibitemOpen
  \bibfield  {author} {\bibinfo {author} {\bibfnamefont {A.~B.}\ \bibnamefont
  {Mundt}}, \bibinfo {author} {\bibfnamefont {A.}~\bibnamefont {Kreuter}},
  \bibinfo {author} {\bibfnamefont {C.}~\bibnamefont {Becher}}, \bibinfo
  {author} {\bibfnamefont {D.}~\bibnamefont {Leibfried}}, \bibinfo {author}
  {\bibfnamefont {J.}~\bibnamefont {Eschner}}, \bibinfo {author} {\bibfnamefont
  {F.}~\bibnamefont {Schmidt-Kaler}},\ and\ \bibinfo {author} {\bibfnamefont
  {R.}~\bibnamefont {Blatt}},\ }\href
  {https://doi.org/10.1103/PhysRevLett.89.103001} {\bibfield  {journal}
  {\bibinfo  {journal} {Phys. Rev. Lett.}\ }\textbf {\bibinfo {volume} {89}},\
  \bibinfo {pages} {103001} (\bibinfo {year} {2002})}\BibitemShut {NoStop}%
\bibitem [{\citenamefont {DeLaubenfels}\ \emph {et~al.}(2015)\citenamefont
  {DeLaubenfels}, \citenamefont {Burkhardt}, \citenamefont {Vittorini},
  \citenamefont {Merrill}, \citenamefont {Brown},\ and\ \citenamefont
  {Amini}}]{delaubenfels2015modulating}%
  \BibitemOpen
  \bibfield  {author} {\bibinfo {author} {\bibfnamefont {T.~E.}\ \bibnamefont
  {DeLaubenfels}}, \bibinfo {author} {\bibfnamefont {K.~A.}\ \bibnamefont
  {Burkhardt}}, \bibinfo {author} {\bibfnamefont {G.}~\bibnamefont
  {Vittorini}}, \bibinfo {author} {\bibfnamefont {J.~T.}\ \bibnamefont
  {Merrill}}, \bibinfo {author} {\bibfnamefont {K.~R.}\ \bibnamefont {Brown}},\
  and\ \bibinfo {author} {\bibfnamefont {J.~M.}\ \bibnamefont {Amini}},\ }\href
  {https://doi.org/10.1103/PhysRevA.92.061402} {\bibfield  {journal} {\bibinfo
  {journal} {Phys. Rev. A}\ }\textbf {\bibinfo {volume} {92}},\ \bibinfo
  {pages} {61402} (\bibinfo {year} {2015})}\BibitemShut {NoStop}%
\bibitem [{\citenamefont {Vasquez}\ \emph {et~al.}(2023)\citenamefont
  {Vasquez}, \citenamefont {Mordini}, \citenamefont {Verni\`ere}, \citenamefont
  {Stadler}, \citenamefont {Malinowski}, \citenamefont {Zhang}, \citenamefont
  {Kienzler}, \citenamefont {Mehta},\ and\ \citenamefont
  {Home}}]{vasquez2022control}%
  \BibitemOpen
  \bibfield  {author} {\bibinfo {author} {\bibfnamefont {A.~R.}\ \bibnamefont
  {Vasquez}}, \bibinfo {author} {\bibfnamefont {C.}~\bibnamefont {Mordini}},
  \bibinfo {author} {\bibfnamefont {C.}~\bibnamefont {Verni\`ere}}, \bibinfo
  {author} {\bibfnamefont {M.}~\bibnamefont {Stadler}}, \bibinfo {author}
  {\bibfnamefont {M.}~\bibnamefont {Malinowski}}, \bibinfo {author}
  {\bibfnamefont {C.}~\bibnamefont {Zhang}}, \bibinfo {author} {\bibfnamefont
  {D.}~\bibnamefont {Kienzler}}, \bibinfo {author} {\bibfnamefont {K.~K.}\
  \bibnamefont {Mehta}},\ and\ \bibinfo {author} {\bibfnamefont {J.~P.}\
  \bibnamefont {Home}},\ }\href
  {https://doi.org/10.1103/PhysRevLett.130.133201} {\bibfield  {journal}
  {\bibinfo  {journal} {Phys. Rev. Lett.}\ }\textbf {\bibinfo {volume} {130}},\
  \bibinfo {pages} {133201} (\bibinfo {year} {2023})}\BibitemShut {NoStop}%
\bibitem [{\citenamefont {Schmiegelow}\ \emph {et~al.}(2016)\citenamefont
  {Schmiegelow}, \citenamefont {Kaufmann}, \citenamefont {Ruster},
  \citenamefont {Schulz}, \citenamefont {Kaushal}, \citenamefont {Hettrich},
  \citenamefont {Schmidt-Kaler},\ and\ \citenamefont
  {Poschinger}}]{schmiegelow2016phasestable}%
  \BibitemOpen
  \bibfield  {author} {\bibinfo {author} {\bibfnamefont {C.~T.}\ \bibnamefont
  {Schmiegelow}}, \bibinfo {author} {\bibfnamefont {H.}~\bibnamefont
  {Kaufmann}}, \bibinfo {author} {\bibfnamefont {T.}~\bibnamefont {Ruster}},
  \bibinfo {author} {\bibfnamefont {J.}~\bibnamefont {Schulz}}, \bibinfo
  {author} {\bibfnamefont {V.}~\bibnamefont {Kaushal}}, \bibinfo {author}
  {\bibfnamefont {M.}~\bibnamefont {Hettrich}}, \bibinfo {author}
  {\bibfnamefont {F.}~\bibnamefont {Schmidt-Kaler}},\ and\ \bibinfo {author}
  {\bibfnamefont {U.~G.}\ \bibnamefont {Poschinger}},\ }\href
  {https://doi.org/10.1103/PhysRevLett.116.033002} {\bibfield  {journal}
  {\bibinfo  {journal} {Phys. Rev. Lett.}\ }\textbf {\bibinfo {volume} {116}},\
  \bibinfo {pages} {33002} (\bibinfo {year} {2016})}\BibitemShut {NoStop}%
\bibitem [{\citenamefont {S\o{}rensen}\ and\ \citenamefont
  {M\o{}lmer}(2000)}]{sorensen2000entanglement}%
  \BibitemOpen
  \bibfield  {author} {\bibinfo {author} {\bibfnamefont {A.}~\bibnamefont
  {S\o{}rensen}}\ and\ \bibinfo {author} {\bibfnamefont {K.}~\bibnamefont
  {M\o{}lmer}},\ }\href {https://doi.org/10.1103/PhysRevA.62.022311} {\bibfield
   {journal} {\bibinfo  {journal} {Phys. Rev. A}\ }\textbf {\bibinfo {volume}
  {62}},\ \bibinfo {pages} {022311} (\bibinfo {year} {2000})}\BibitemShut
  {NoStop}%
\bibitem [{\citenamefont {Mehta}\ \emph {et~al.}(2019)\citenamefont {Mehta},
  \citenamefont {Zhang}, \citenamefont {Miller},\ and\ \citenamefont
  {Home}}]{mehta2019towards}%
  \BibitemOpen
  \bibfield  {author} {\bibinfo {author} {\bibfnamefont {K.~K.}\ \bibnamefont
  {Mehta}}, \bibinfo {author} {\bibfnamefont {C.}~\bibnamefont {Zhang}},
  \bibinfo {author} {\bibfnamefont {S.}~\bibnamefont {Miller}},\ and\ \bibinfo
  {author} {\bibfnamefont {J.~P.}\ \bibnamefont {Home}},\ }in\ \href
  {https://doi.org/10.1117/12.2507647} {\emph {\bibinfo {booktitle} {Proc.
  SPIE}}},\ Vol.\ \bibinfo {volume} {10933}\ (\bibinfo {year} {2019})\ p.\
  \bibinfo {pages} {109330B}\BibitemShut {NoStop}%
\bibitem [{\citenamefont {Kielpinski}\ \emph {et~al.}(2002)\citenamefont
  {Kielpinski}, \citenamefont {Monroe},\ and\ \citenamefont
  {Wineland}}]{Kielpinski2002architecture}%
  \BibitemOpen
  \bibfield  {author} {\bibinfo {author} {\bibfnamefont {D.}~\bibnamefont
  {Kielpinski}}, \bibinfo {author} {\bibfnamefont {C.}~\bibnamefont {Monroe}},\
  and\ \bibinfo {author} {\bibfnamefont {D.~J.}\ \bibnamefont {Wineland}},\
  }\href {https://doi.org/10.1038/nature00784} {\bibfield  {journal} {\bibinfo
  {journal} {Nature}\ }\textbf {\bibinfo {volume} {417}},\ \bibinfo {pages}
  {709} (\bibinfo {year} {2002})}\BibitemShut {NoStop}%
\bibitem [{\citenamefont {Sch{\"{a}}fer}\ \emph {et~al.}(2018)\citenamefont
  {Sch{\"{a}}fer}, \citenamefont {Ballance}, \citenamefont {Thirumalai},
  \citenamefont {Stephenson}, \citenamefont {Ballance}, \citenamefont
  {Steane},\ and\ \citenamefont {Lucas}}]{schafer2018fast}%
  \BibitemOpen
  \bibfield  {author} {\bibinfo {author} {\bibfnamefont {V.~M.}\ \bibnamefont
  {Sch{\"{a}}fer}}, \bibinfo {author} {\bibfnamefont {C.~J.}\ \bibnamefont
  {Ballance}}, \bibinfo {author} {\bibfnamefont {K.}~\bibnamefont
  {Thirumalai}}, \bibinfo {author} {\bibfnamefont {L.~J.}\ \bibnamefont
  {Stephenson}}, \bibinfo {author} {\bibfnamefont {T.~G.}\ \bibnamefont
  {Ballance}}, \bibinfo {author} {\bibfnamefont {A.~M.}\ \bibnamefont
  {Steane}},\ and\ \bibinfo {author} {\bibfnamefont {D.~M.}\ \bibnamefont
  {Lucas}},\ }\href {https://doi.org/10.1038/nature25737} {\bibfield  {journal}
  {\bibinfo  {journal} {Nature}\ }\textbf {\bibinfo {volume} {555}},\ \bibinfo
  {pages} {75} (\bibinfo {year} {2018})}\BibitemShut {NoStop}%
\bibitem [{\citenamefont {Reznik}\ \emph {et~al.}(2005)\citenamefont {Reznik},
  \citenamefont {Retzker},\ and\ \citenamefont {Silman}}]{reznik2005violating}%
  \BibitemOpen
  \bibfield  {author} {\bibinfo {author} {\bibfnamefont {B.}~\bibnamefont
  {Reznik}}, \bibinfo {author} {\bibfnamefont {A.}~\bibnamefont {Retzker}},\
  and\ \bibinfo {author} {\bibfnamefont {J.}~\bibnamefont {Silman}},\ }\href
  {https://doi.org/10.1103/PhysRevA.71.042104} {\bibfield  {journal} {\bibinfo
  {journal} {Phys. Rev. A}\ }\textbf {\bibinfo {volume} {71}},\ \bibinfo
  {pages} {042104} (\bibinfo {year} {2005})}\BibitemShut {NoStop}%
\bibitem [{\citenamefont {Retzker}\ \emph {et~al.}(2005)\citenamefont
  {Retzker}, \citenamefont {Cirac},\ and\ \citenamefont
  {Reznik}}]{retzker2005detecting}%
  \BibitemOpen
  \bibfield  {author} {\bibinfo {author} {\bibfnamefont {A.}~\bibnamefont
  {Retzker}}, \bibinfo {author} {\bibfnamefont {J.~I.}\ \bibnamefont {Cirac}},\
  and\ \bibinfo {author} {\bibfnamefont {B.}~\bibnamefont {Reznik}},\ }\href
  {https://doi.org/10.1103/PhysRevLett.94.050504} {\bibfield  {journal}
  {\bibinfo  {journal} {Phys. Rev. Lett.}\ }\textbf {\bibinfo {volume} {94}},\
  \bibinfo {pages} {050504} (\bibinfo {year} {2005})}\BibitemShut {NoStop}%
\bibitem [{\citenamefont {Drechsler}\ \emph {et~al.}(2021)\citenamefont
  {Drechsler}, \citenamefont {Wolf}, \citenamefont {Schmiegelow},\ and\
  \citenamefont {Schmidt-Kaler}}]{drechsler2021optical}%
  \BibitemOpen
  \bibfield  {author} {\bibinfo {author} {\bibfnamefont {M.}~\bibnamefont
  {Drechsler}}, \bibinfo {author} {\bibfnamefont {S.}~\bibnamefont {Wolf}},
  \bibinfo {author} {\bibfnamefont {C.~T.}\ \bibnamefont {Schmiegelow}},\ and\
  \bibinfo {author} {\bibfnamefont {F.}~\bibnamefont {Schmidt-Kaler}},\ }\href
  {https://doi.org/10.1103/PhysRevLett.127.143602} {\bibfield  {journal}
  {\bibinfo  {journal} {Phys. Rev. Lett.}\ }\textbf {\bibinfo {volume} {127}},\
  \bibinfo {pages} {143602} (\bibinfo {year} {2021})}\BibitemShut {NoStop}%
\bibitem [{\citenamefont {Qian}\ \emph {et~al.}(2021)\citenamefont {Qian},
  \citenamefont {Cui}, \citenamefont {Luo}, \citenamefont {Zheng},
  \citenamefont {Huang}, \citenamefont {Ai}, \citenamefont {He}, \citenamefont
  {Li},\ and\ \citenamefont {Guo}}]{qian2021super}%
  \BibitemOpen
  \bibfield  {author} {\bibinfo {author} {\bibfnamefont {Z.-H.}\ \bibnamefont
  {Qian}}, \bibinfo {author} {\bibfnamefont {J.-M.}\ \bibnamefont {Cui}},
  \bibinfo {author} {\bibfnamefont {X.-W.}\ \bibnamefont {Luo}}, \bibinfo
  {author} {\bibfnamefont {Y.-X.}\ \bibnamefont {Zheng}}, \bibinfo {author}
  {\bibfnamefont {Y.-F.}\ \bibnamefont {Huang}}, \bibinfo {author}
  {\bibfnamefont {M.-Z.}\ \bibnamefont {Ai}}, \bibinfo {author} {\bibfnamefont
  {R.}~\bibnamefont {He}}, \bibinfo {author} {\bibfnamefont {C.-F.}\
  \bibnamefont {Li}},\ and\ \bibinfo {author} {\bibfnamefont {G.-C.}\
  \bibnamefont {Guo}},\ }\href {https://doi.org/10.1103/PhysRevLett.127.263603}
  {\bibfield  {journal} {\bibinfo  {journal} {Phys. Rev. Lett.}\ }\textbf
  {\bibinfo {volume} {127}},\ \bibinfo {pages} {263603} (\bibinfo {year}
  {2021})}\BibitemShut {NoStop}%
\bibitem [{\citenamefont {Yudin}\ \emph {et~al.}(2010)\citenamefont {Yudin},
  \citenamefont {Taichenachev}, \citenamefont {Oates}, \citenamefont {Barber},
  \citenamefont {Lemke}, \citenamefont {Ludlow}, \citenamefont {Sterr},
  \citenamefont {Lisdat},\ and\ \citenamefont {Riehle}}]{yudin2010hyper}%
  \BibitemOpen
  \bibfield  {author} {\bibinfo {author} {\bibfnamefont {V.~I.}\ \bibnamefont
  {Yudin}}, \bibinfo {author} {\bibfnamefont {A.~V.}\ \bibnamefont
  {Taichenachev}}, \bibinfo {author} {\bibfnamefont {C.~W.}\ \bibnamefont
  {Oates}}, \bibinfo {author} {\bibfnamefont {Z.~W.}\ \bibnamefont {Barber}},
  \bibinfo {author} {\bibfnamefont {N.~D.}\ \bibnamefont {Lemke}}, \bibinfo
  {author} {\bibfnamefont {A.~D.}\ \bibnamefont {Ludlow}}, \bibinfo {author}
  {\bibfnamefont {U.}~\bibnamefont {Sterr}}, \bibinfo {author} {\bibfnamefont
  {C.}~\bibnamefont {Lisdat}},\ and\ \bibinfo {author} {\bibfnamefont
  {F.}~\bibnamefont {Riehle}},\ }\href
  {https://doi.org/10.1103/PhysRevA.82.011804} {\bibfield  {journal} {\bibinfo
  {journal} {Phys. Rev. A}\ }\textbf {\bibinfo {volume} {82}},\ \bibinfo
  {pages} {011804} (\bibinfo {year} {2010})}\BibitemShut {NoStop}%
\bibitem [{\citenamefont {Vetlugin}\ \emph {et~al.}(2022)\citenamefont
  {Vetlugin}, \citenamefont {Guo}, \citenamefont {Soci},\ and\ \citenamefont
  {Zheludev}}]{vetlugin2022deterministic}%
  \BibitemOpen
  \bibfield  {author} {\bibinfo {author} {\bibfnamefont {A.~N.}\ \bibnamefont
  {Vetlugin}}, \bibinfo {author} {\bibfnamefont {R.}~\bibnamefont {Guo}},
  \bibinfo {author} {\bibfnamefont {C.}~\bibnamefont {Soci}},\ and\ \bibinfo
  {author} {\bibfnamefont {N.~I.}\ \bibnamefont {Zheludev}},\ }\href
  {https://doi.org/10.1103/PhysRevA.106.012402} {\bibfield  {journal} {\bibinfo
   {journal} {Phys. Rev. A}\ }\textbf {\bibinfo {volume} {106}},\ \bibinfo
  {pages} {012402} (\bibinfo {year} {2022})}\BibitemShut {NoStop}%
\bibitem [{\citenamefont {Cirac}\ \emph {et~al.}(1993)\citenamefont {Cirac},
  \citenamefont {Blatt}, \citenamefont {Parkins},\ and\ \citenamefont
  {Zoller}}]{cirac1993preparation}%
  \BibitemOpen
  \bibfield  {author} {\bibinfo {author} {\bibfnamefont {J.~I.}\ \bibnamefont
  {Cirac}}, \bibinfo {author} {\bibfnamefont {R.}~\bibnamefont {Blatt}},
  \bibinfo {author} {\bibfnamefont {A.~S.}\ \bibnamefont {Parkins}},\ and\
  \bibinfo {author} {\bibfnamefont {P.}~\bibnamefont {Zoller}},\ }\href
  {https://doi.org/10.1103/PhysRevLett.70.762} {\bibfield  {journal} {\bibinfo
  {journal} {Phys. Rev. Lett.}\ }\textbf {\bibinfo {volume} {70}},\ \bibinfo
  {pages} {762} (\bibinfo {year} {1993})}\BibitemShut {NoStop}%
\bibitem [{\citenamefont {Cirac}\ \emph {et~al.}(1994)\citenamefont {Cirac},
  \citenamefont {Blatt}, \citenamefont {Parkins},\ and\ \citenamefont
  {Zoller}}]{cirac1994quantum}%
  \BibitemOpen
  \bibfield  {author} {\bibinfo {author} {\bibfnamefont {J.~I.}\ \bibnamefont
  {Cirac}}, \bibinfo {author} {\bibfnamefont {R.}~\bibnamefont {Blatt}},
  \bibinfo {author} {\bibfnamefont {A.~S.}\ \bibnamefont {Parkins}},\ and\
  \bibinfo {author} {\bibfnamefont {P.}~\bibnamefont {Zoller}},\ }\href
  {https://doi.org/10.1103/PhysRevA.49.1202} {\bibfield  {journal} {\bibinfo
  {journal} {Phys. Rev. A}\ }\textbf {\bibinfo {volume} {49}},\ \bibinfo
  {pages} {1202} (\bibinfo {year} {1994})}\BibitemShut {NoStop}%
\bibitem [{\citenamefont {Wu}\ and\ \citenamefont {Yang}(1997)}]{wu1997jaynes}%
  \BibitemOpen
  \bibfield  {author} {\bibinfo {author} {\bibfnamefont {Y.}~\bibnamefont
  {Wu}}\ and\ \bibinfo {author} {\bibfnamefont {X.}~\bibnamefont {Yang}},\
  }\href {https://doi.org/10.1103/PhysRevLett.78.3086} {\bibfield  {journal}
  {\bibinfo  {journal} {Phys. Rev. Lett.}\ }\textbf {\bibinfo {volume} {78}},\
  \bibinfo {pages} {3086} (\bibinfo {year} {1997})}\BibitemShut {NoStop}%
\bibitem [{Note1()}]{Note1}%
  \BibitemOpen
  \bibinfo {note} {$\protect \hat {\sigma }_{\alpha }^{(i)} =\protect
  \underbrace {\protect \hat {\protect \mathbb {I}}_2 \otimes ...\otimes
  \protect \hat {\protect \mathbb {I}}_2}_{i-1} \otimes \protect \hat {\sigma
  }_{\alpha }\otimes \protect \underbrace {\protect \hat {\protect \mathbb
  {I}}_2 \otimes ...\otimes \protect \hat {\protect \mathbb {I}}_2}_{n-i}$,
  where $\alpha \in \{+, -, x, y\}$ in this work}\BibitemShut {NoStop}%
\bibitem [{sup()}]{supplementary}%
  \BibitemOpen
  \href@noop {} {}\bibinfo {note} {See Supplemental Material for additional
  steps in the derivation of the equations presented in the text and for
  experimental details.}\BibitemShut {Stop}%
\bibitem [{\citenamefont {Roos}(2008)}]{Roos2008ion}%
  \BibitemOpen
  \bibfield  {author} {\bibinfo {author} {\bibfnamefont {C.~F.}\ \bibnamefont
  {Roos}},\ }\href {https://doi.org/10.1088/1367-2630/10/1/013002} {\bibfield
  {journal} {\bibinfo  {journal} {New J. Phys.}\ }\textbf {\bibinfo {volume}
  {10}},\ \bibinfo {pages} {013002} (\bibinfo {year} {2008})}\BibitemShut
  {NoStop}%
\bibitem [{\citenamefont {Sutherland}\ \emph {et~al.}(2019)\citenamefont
  {Sutherland}, \citenamefont {Srinivas}, \citenamefont {Burd}, \citenamefont
  {Leibfried}, \citenamefont {Wilson}, \citenamefont {Wineland}, \citenamefont
  {Allcock}, \citenamefont {Slichter},\ and\ \citenamefont
  {Libby}}]{sutherland2019versatile}%
  \BibitemOpen
  \bibfield  {author} {\bibinfo {author} {\bibfnamefont {R.}~\bibnamefont
  {Sutherland}}, \bibinfo {author} {\bibfnamefont {R.}~\bibnamefont
  {Srinivas}}, \bibinfo {author} {\bibfnamefont {S.~C.}\ \bibnamefont {Burd}},
  \bibinfo {author} {\bibfnamefont {D.}~\bibnamefont {Leibfried}}, \bibinfo
  {author} {\bibfnamefont {A.~C.}\ \bibnamefont {Wilson}}, \bibinfo {author}
  {\bibfnamefont {D.~J.}\ \bibnamefont {Wineland}}, \bibinfo {author}
  {\bibfnamefont {D.}~\bibnamefont {Allcock}}, \bibinfo {author} {\bibfnamefont
  {D.}~\bibnamefont {Slichter}},\ and\ \bibinfo {author} {\bibfnamefont
  {S.}~\bibnamefont {Libby}},\ }\href
  {https://doi.org/10.1088/1367-2630/ab0be5} {\bibfield  {journal} {\bibinfo
  {journal} {New J. Phys.}\ }\textbf {\bibinfo {volume} {21}},\ \bibinfo
  {pages} {033033} (\bibinfo {year} {2019})}\BibitemShut {NoStop}%
\bibitem [{\citenamefont {B\ifmmode \u{a}\else \u{a}\fi{}z\ifmmode~\u{a}\else
  \u{a}\fi{}van}\ \emph {et~al.}(2023)\citenamefont {B\ifmmode \u{a}\else
  \u{a}\fi{}z\ifmmode~\u{a}\else \u{a}\fi{}van}, \citenamefont {Saner},
  \citenamefont {Minder}, \citenamefont {Hughes}, \citenamefont {Sutherland},
  \citenamefont {Lucas}, \citenamefont {Srinivas},\ and\ \citenamefont
  {Ballance}}]{Bazavan2022synthesizing}%
  \BibitemOpen
  \bibfield  {author} {\bibinfo {author} {\bibfnamefont {O.}~\bibnamefont
  {B\ifmmode \u{a}\else \u{a}\fi{}z\ifmmode~\u{a}\else \u{a}\fi{}van}},
  \bibinfo {author} {\bibfnamefont {S.}~\bibnamefont {Saner}}, \bibinfo
  {author} {\bibfnamefont {M.}~\bibnamefont {Minder}}, \bibinfo {author}
  {\bibfnamefont {A.~C.}\ \bibnamefont {Hughes}}, \bibinfo {author}
  {\bibfnamefont {R.~T.}\ \bibnamefont {Sutherland}}, \bibinfo {author}
  {\bibfnamefont {D.~M.}\ \bibnamefont {Lucas}}, \bibinfo {author}
  {\bibfnamefont {R.}~\bibnamefont {Srinivas}},\ and\ \bibinfo {author}
  {\bibfnamefont {C.~J.}\ \bibnamefont {Ballance}},\ }\href
  {https://doi.org/10.1103/PhysRevA.107.022617} {\bibfield  {journal} {\bibinfo
   {journal} {Phys. Rev. A}\ }\textbf {\bibinfo {volume} {107}},\ \bibinfo
  {pages} {022617} (\bibinfo {year} {2023})}\BibitemShut {NoStop}%
\bibitem [{\citenamefont {Thirumalai}(2019)}]{thirumalai2019high}%
  \BibitemOpen
  \bibfield  {author} {\bibinfo {author} {\bibfnamefont {K.}~\bibnamefont
  {Thirumalai}},\ }\emph {\bibinfo {title} {{High-fidelity mixed species
  entanglement of trapped ions}}},\ \href@noop {} {Ph.D. thesis},\ \bibinfo
  {school} {University of Oxford} (\bibinfo {year} {2019})\BibitemShut
  {NoStop}%
\bibitem [{\citenamefont {Sch{\"a}fer}(2018)}]{schafer2018fastthesis}%
  \BibitemOpen
  \bibfield  {author} {\bibinfo {author} {\bibfnamefont {V.}~\bibnamefont
  {Sch{\"a}fer}},\ }\emph {\bibinfo {title} {Fast gates and mixed-species
  entanglement with trapped ions}},\ \href@noop {} {Ph.D. thesis},\ \bibinfo
  {school} {University of Oxford} (\bibinfo {year} {2018})\BibitemShut
  {NoStop}%
\bibitem [{\citenamefont {Knill}\ \emph {et~al.}(2008)\citenamefont {Knill},
  \citenamefont {Leibfried}, \citenamefont {Reichle}, \citenamefont {Britton},
  \citenamefont {Blakestad}, \citenamefont {Jost}, \citenamefont {Langer},
  \citenamefont {Ozeri}, \citenamefont {Seidelin},\ and\ \citenamefont
  {Wineland}}]{knill2008randomized}%
  \BibitemOpen
  \bibfield  {author} {\bibinfo {author} {\bibfnamefont {E.}~\bibnamefont
  {Knill}}, \bibinfo {author} {\bibfnamefont {D.}~\bibnamefont {Leibfried}},
  \bibinfo {author} {\bibfnamefont {R.}~\bibnamefont {Reichle}}, \bibinfo
  {author} {\bibfnamefont {J.}~\bibnamefont {Britton}}, \bibinfo {author}
  {\bibfnamefont {R.~B.}\ \bibnamefont {Blakestad}}, \bibinfo {author}
  {\bibfnamefont {J.~D.}\ \bibnamefont {Jost}}, \bibinfo {author}
  {\bibfnamefont {C.}~\bibnamefont {Langer}}, \bibinfo {author} {\bibfnamefont
  {R.}~\bibnamefont {Ozeri}}, \bibinfo {author} {\bibfnamefont
  {S.}~\bibnamefont {Seidelin}},\ and\ \bibinfo {author} {\bibfnamefont
  {D.~J.}\ \bibnamefont {Wineland}},\ }\href
  {https://doi.org/10.1103/PhysRevA.77.012307} {\bibfield  {journal} {\bibinfo
  {journal} {Phys. Rev. A}\ }\textbf {\bibinfo {volume} {77}},\ \bibinfo
  {pages} {012307} (\bibinfo {year} {2008})}\BibitemShut {NoStop}%
\bibitem [{fnp()}]{fnpulseshape}%
  \BibitemOpen
  \href@noop {} {}\bibinfo {note} {The ramp shape is a $\sin(\pi t/2t_R)^2$
  with a total rise time given by the ramp duration $t_R$.}\BibitemShut {Stop}%
\bibitem [{fni()}]{fninterference}%
  \BibitemOpen
  \href@noop {} {}\bibinfo {note} {Note that $\Omega$ in $\Omega_{\rm
  SDF}(\Omega, \delta)$ for the SW has been rescaled by a factor of $2$ to
  account for the power enhancement due to interference, and in order to
  compare fairly the TW to the SW.}\BibitemShut {Stop}%
\bibitem [{\citenamefont {Steane}\ \emph {et~al.}(2014)\citenamefont {Steane},
  \citenamefont {Imreh}, \citenamefont {Home},\ and\ \citenamefont
  {Leibfried}}]{Steane2014pulsed}%
  \BibitemOpen
  \bibfield  {author} {\bibinfo {author} {\bibfnamefont {A.~M.}\ \bibnamefont
  {Steane}}, \bibinfo {author} {\bibfnamefont {G.}~\bibnamefont {Imreh}},
  \bibinfo {author} {\bibfnamefont {J.~P.}\ \bibnamefont {Home}},\ and\
  \bibinfo {author} {\bibfnamefont {D.}~\bibnamefont {Leibfried}},\ }\href
  {https://doi.org/10.1088/1367-2630/16/5/053049} {\bibfield  {journal}
  {\bibinfo  {journal} {New J. Phys.}\ }\textbf {\bibinfo {volume} {16}},\
  \bibinfo {pages} {53049} (\bibinfo {year} {2014})}\BibitemShut {NoStop}%
\bibitem [{\citenamefont {Palmero}\ \emph {et~al.}(2017)\citenamefont
  {Palmero}, \citenamefont {Mart\'{i}nez-Garaot}, \citenamefont {Leibfried},
  \citenamefont {Wineland},\ and\ \citenamefont {Muga}}]{palermo2017fast}%
  \BibitemOpen
  \bibfield  {author} {\bibinfo {author} {\bibfnamefont {M.}~\bibnamefont
  {Palmero}}, \bibinfo {author} {\bibfnamefont {S.}~\bibnamefont
  {Mart\'{i}nez-Garaot}}, \bibinfo {author} {\bibfnamefont {D.}~\bibnamefont
  {Leibfried}}, \bibinfo {author} {\bibfnamefont {D.~J.}\ \bibnamefont
  {Wineland}},\ and\ \bibinfo {author} {\bibfnamefont {J.~G.}\ \bibnamefont
  {Muga}},\ }\href {https://doi.org/10.1103/PhysRevA.95.022328} {\bibfield
  {journal} {\bibinfo  {journal} {Phys. Rev. A}\ }\textbf {\bibinfo {volume}
  {95}},\ \bibinfo {pages} {22328} (\bibinfo {year} {2017})}\BibitemShut
  {NoStop}%
\bibitem [{Note2()}]{Note2}%
  \BibitemOpen
  \bibinfo {note} {The effective gate duration is approximately equal to the
  full-width half maximum (FWHM) of the pulse shape. The pulse duration from
  start to end is $2\pi /\delta _g + t_R$, where $t_R$ is the ramp duration.
  The ramp shape is as defined in \cite {fnpulseshape}.}\BibitemShut {Stop}%
\bibitem [{\citenamefont {Wang}\ \emph {et~al.}(2011)\citenamefont {Wang},
  \citenamefont {{Hao Low}}, \citenamefont {Lachenmyer}, \citenamefont {Ge},
  \citenamefont {Herskind},\ and\ \citenamefont {Chuang}}]{wang2011laser}%
  \BibitemOpen
  \bibfield  {author} {\bibinfo {author} {\bibfnamefont {S.~X.}\ \bibnamefont
  {Wang}}, \bibinfo {author} {\bibfnamefont {G.}~\bibnamefont {{Hao Low}}},
  \bibinfo {author} {\bibfnamefont {N.~S.}\ \bibnamefont {Lachenmyer}},
  \bibinfo {author} {\bibfnamefont {Y.}~\bibnamefont {Ge}}, \bibinfo {author}
  {\bibfnamefont {P.~F.}\ \bibnamefont {Herskind}},\ and\ \bibinfo {author}
  {\bibfnamefont {I.~L.}\ \bibnamefont {Chuang}},\ }\href
  {https://doi.org/https://doi.org/10.1063/1.3662118} {\bibfield  {journal}
  {\bibinfo  {journal} {J. Appl. Phys.}\ }\textbf {\bibinfo {volume} {110}},\
  \bibinfo {pages} {104901} (\bibinfo {year} {2011})}\BibitemShut {NoStop}%
\bibitem [{\citenamefont {Niffenegger}\ \emph {et~al.}(2020)\citenamefont
  {Niffenegger}, \citenamefont {Stuart}, \citenamefont {Sorace-Agaskar},
  \citenamefont {Kharas}, \citenamefont {Bramhavar}, \citenamefont {Bruzewicz},
  \citenamefont {Loh}, \citenamefont {Maxson}, \citenamefont {McConnell},
  \citenamefont {Reens}, \citenamefont {West}, \citenamefont {Sage},\ and\
  \citenamefont {Chiaverini}}]{niffenegger2020integrated}%
  \BibitemOpen
  \bibfield  {author} {\bibinfo {author} {\bibfnamefont {R.~J.}\ \bibnamefont
  {Niffenegger}}, \bibinfo {author} {\bibfnamefont {J.}~\bibnamefont {Stuart}},
  \bibinfo {author} {\bibfnamefont {C.}~\bibnamefont {Sorace-Agaskar}},
  \bibinfo {author} {\bibfnamefont {D.}~\bibnamefont {Kharas}}, \bibinfo
  {author} {\bibfnamefont {S.}~\bibnamefont {Bramhavar}}, \bibinfo {author}
  {\bibfnamefont {C.~D.}\ \bibnamefont {Bruzewicz}}, \bibinfo {author}
  {\bibfnamefont {W.}~\bibnamefont {Loh}}, \bibinfo {author} {\bibfnamefont
  {R.~T.}\ \bibnamefont {Maxson}}, \bibinfo {author} {\bibfnamefont
  {R.}~\bibnamefont {McConnell}}, \bibinfo {author} {\bibfnamefont
  {D.}~\bibnamefont {Reens}}, \bibinfo {author} {\bibfnamefont {G.~N.}\
  \bibnamefont {West}}, \bibinfo {author} {\bibfnamefont {J.~M.}\ \bibnamefont
  {Sage}},\ and\ \bibinfo {author} {\bibfnamefont {J.}~\bibnamefont
  {Chiaverini}},\ }\href {https://doi.org/10.1038/s41586-020-2811-x} {\bibfield
   {journal} {\bibinfo  {journal} {Nature}\ }\textbf {\bibinfo {volume}
  {586}},\ \bibinfo {pages} {538} (\bibinfo {year} {2020})}\BibitemShut
  {NoStop}%
\bibitem [{\citenamefont {Mehta}\ \emph {et~al.}(2020)\citenamefont {Mehta},
  \citenamefont {Zhang}, \citenamefont {Malinowski}, \citenamefont {Nguyen},
  \citenamefont {Stadler},\ and\ \citenamefont {Home}}]{Mehta2020}%
  \BibitemOpen
  \bibfield  {author} {\bibinfo {author} {\bibfnamefont {K.~K.}\ \bibnamefont
  {Mehta}}, \bibinfo {author} {\bibfnamefont {C.}~\bibnamefont {Zhang}},
  \bibinfo {author} {\bibfnamefont {M.}~\bibnamefont {Malinowski}}, \bibinfo
  {author} {\bibfnamefont {T.-L.}\ \bibnamefont {Nguyen}}, \bibinfo {author}
  {\bibfnamefont {M.}~\bibnamefont {Stadler}},\ and\ \bibinfo {author}
  {\bibfnamefont {J.~P.}\ \bibnamefont {Home}},\ }\href
  {https://doi.org/10.1038/s41586-020-2823-6} {\bibfield  {journal} {\bibinfo
  {journal} {Nature}\ }\textbf {\bibinfo {volume} {586}},\ \bibinfo {pages}
  {533} (\bibinfo {year} {2020})}\BibitemShut {NoStop}%
\bibitem [{\citenamefont {Garc\'{\i}a-Ripoll}\ \emph
  {et~al.}(2003)\citenamefont {Garc\'{\i}a-Ripoll}, \citenamefont {Zoller},\
  and\ \citenamefont {Cirac}}]{garcia2003speed}%
  \BibitemOpen
  \bibfield  {author} {\bibinfo {author} {\bibfnamefont {J.~J.}\ \bibnamefont
  {Garc\'{\i}a-Ripoll}}, \bibinfo {author} {\bibfnamefont {P.}~\bibnamefont
  {Zoller}},\ and\ \bibinfo {author} {\bibfnamefont {J.~I.}\ \bibnamefont
  {Cirac}},\ }\href {https://doi.org/10.1103/PhysRevLett.91.157901} {\bibfield
  {journal} {\bibinfo  {journal} {Phys. Rev. Lett.}\ }\textbf {\bibinfo
  {volume} {91}},\ \bibinfo {pages} {157901} (\bibinfo {year}
  {2003})}\BibitemShut {NoStop}%
\bibitem [{\citenamefont {Wong-Campos}\ \emph {et~al.}(2017)\citenamefont
  {Wong-Campos}, \citenamefont {Moses}, \citenamefont {Johnson},\ and\
  \citenamefont {Monroe}}]{wong2017demonstration}%
  \BibitemOpen
  \bibfield  {author} {\bibinfo {author} {\bibfnamefont {J.~D.}\ \bibnamefont
  {Wong-Campos}}, \bibinfo {author} {\bibfnamefont {S.~A.}\ \bibnamefont
  {Moses}}, \bibinfo {author} {\bibfnamefont {K.~G.}\ \bibnamefont {Johnson}},\
  and\ \bibinfo {author} {\bibfnamefont {C.}~\bibnamefont {Monroe}},\ }\href
  {https://doi.org/10.1103/PhysRevLett.119.230501} {\bibfield  {journal}
  {\bibinfo  {journal} {Phys. Rev. Lett.}\ }\textbf {\bibinfo {volume} {119}},\
  \bibinfo {pages} {230501} (\bibinfo {year} {2017})}\BibitemShut {NoStop}%
\bibitem [{\citenamefont {Zhang}\ \emph {et~al.}(2020)\citenamefont {Zhang},
  \citenamefont {Pokorny}, \citenamefont {Li}, \citenamefont {Higgins},
  \citenamefont {P{\"{o}}schl}, \citenamefont {Lesanovsky},\ and\ \citenamefont
  {Hennrich}}]{zhang2020submicrosecond}%
  \BibitemOpen
  \bibfield  {author} {\bibinfo {author} {\bibfnamefont {C.}~\bibnamefont
  {Zhang}}, \bibinfo {author} {\bibfnamefont {F.}~\bibnamefont {Pokorny}},
  \bibinfo {author} {\bibfnamefont {W.}~\bibnamefont {Li}}, \bibinfo {author}
  {\bibfnamefont {G.}~\bibnamefont {Higgins}}, \bibinfo {author} {\bibfnamefont
  {A.}~\bibnamefont {P{\"{o}}schl}}, \bibinfo {author} {\bibfnamefont
  {I.}~\bibnamefont {Lesanovsky}},\ and\ \bibinfo {author} {\bibfnamefont
  {M.}~\bibnamefont {Hennrich}},\ }\href
  {https://doi.org/10.1038/s41586-020-2152-9} {\bibfield  {journal} {\bibinfo
  {journal} {Nature}\ }\textbf {\bibinfo {volume} {580}},\ \bibinfo {pages}
  {345} (\bibinfo {year} {2020})}\BibitemShut {NoStop}%
\end{thebibliography}%
\renewcommand{\thefigure}{B.\arabic{figure}}
\setcounter{figure}{0}
\onecolumngrid
\begin{center}
\vspace{5 mm}
\textbf{\large Supplemental Material for:\\
Breaking the entangling gate speed limit for trapped-ion qubits using a phase-stable standing wave}\\
\end{center}
\twocolumngrid
\section{Derivation of the standing wave interaction}
We present the additional steps required to reach Eq.~\eqref{eq:monochrmatic_sw} \cite{cirac1993preparation,cirac1994quantum,wu1997jaynes}. We consider the interaction of a traveling monochromatic field interacting with a string of ions,
\begin{equation}
    \hat{H}_{\mathrm{TW}} = \frac{\hbar\Omega}{2} e^{i(\phi_1 + \eta (\hat{a}e^{-i \omega_z t}+\hat{a}^\dagger e^{i \omega_z t})-\delta t)}\hat{S}_{+} + \mathrm{h.c.} 
\end{equation}
This expression is in the interaction picture w.r.t. the qubit frequency $\omega_0$, and the motional mode frequency $\omega_z$, after the rotating wave approximation
w.r.t. $\omega_0$.
For the derivation of the SW we further assume that the ions are spaced such that the optical phase differs by an integer multiple of $\pi$ at each of the ions.
We add a second traveling monochromatic field counter-propagating to the first with $\phi_2$, $\eta_2 = \bm{k_2} \bm{\hat{z}}= - \bm{k} \bm{\hat{z}} = -\eta$ and $\Omega_2 \propto Q_{ij} \partial^i E_2^j = - Q_{ij} \partial^i E_1^j  \propto -\Omega$ where $Q_{ij}$ is the quadrupole tensor,
\begin{align}
    \hat{H}_{\mathrm{SW}} &= \hat{H}_{\mathrm{TW}}^1+ \hat{H}_{\mathrm{TW}}^2\\
    \begin{split}
    &=\frac{\hbar\Omega}{2} e^{i(\phi_1 + \eta (\hat{a}e^{-i \omega_z t}+\hat{a}^\dagger e^{i \omega_z t})-\delta t)}\hat{S}_{+} \\
    &- \frac{\hbar\Omega}{2} e^{i(\phi_2 - \eta (\hat{a}e^{-i \omega_z t}+\hat{a}^\dagger e^{i \omega_z t})-\delta t)}\hat{S}_{+} 
    + \mathrm{h.c.}
    \end{split}\\
    \begin{split}
    &= i \hbar\Omega\ e^{i\left(\frac{\phi_1+\phi_2}{2} -\delta t\right)}\sin\left(\eta (\hat{a}e^{-i \omega_z t}+\hat{a}^\dagger e^{i \omega_z t})+\frac{\phi_1-\phi_2}{2}\right)\hat{S}_+\\
    &+ \mathrm{h.c.}
    \end{split}
\end{align}
The sideband and carrier coupling can be separated using $\sin(x+y) = \cos(x)\sin(y)+ \sin(x)\cos(y)$:
\begin{align}
    \begin{split}
    \hat{H}_{\rm SW}&= e^{-i\delta t}\hbar\Omega\ e^{i\tilde{\phi}} \hat{S}_{+} \sin\left(\eta (\hat{a}e^{-i \omega_z t} + \hat{a}^\dagger e^{i \omega_z t})\right)\cos\left(\frac{\Delta\phi}{2} \right) \\
    & + e^{-i\delta t}\hbar\Omega\ e^{i\tilde{\phi}}\hat{S}_{+} \cos\left(\eta (\hat{a}e^{-i \omega_z t} + \hat{a}^\dagger e^{i \omega_z t})\right)\sin\left(\frac{\Delta\phi}{2} \right) \\
    & +\mathrm{h.c.}, 
    \end{split}
\end{align}
where we absorb the factor of $i$ in the mean phase $\tilde{\phi} = (\phi_1+\phi_2 + \pi)/2$ and define the difference phase $\Delta\phi = (\phi_1-\phi_2)$. These conventions were also used in the main text.
We note that in the vicinity of $\Delta\phi=0$, the interaction strength of the carrier coupling depends linearly on $\Delta\phi$, while around $\Delta\phi=\pi$ the dependence is quadratic. This leads to a quadratic resp.~quartic dependence in transfer probability as discussed in the main text and observed in Fig.~\ref{fig:lattice-single-rotations}(a).

Using the Lamb-Dicke expansion $\eta \ll \pi$, we obtain
\begin{equation}
\begin{split}
    \hat{H}_\mathrm{SW} &= e^{-i\delta t}\hbar\eta\Omega\ e^{i\tilde{\phi}} \hat{S}_{+} (\hat{a}e^{-i \omega_z t} + \hat{a}^\dagger e^{i \omega_z t}) \cos\left(\frac{\Delta\phi}{2}\right) \\
    &+ e^{ -i\delta t}\hbar\Omega\ e^{i\tilde{\phi}}\hat{S}_{+} \sin\left(\frac{\Delta\phi}{2}\right) \\
    &+\cos\left(\frac{\Delta\phi}{2}\right) \mathcal{O}\left(\eta^{2j+1}\right)+\sin\left(\frac{\Delta\phi}{2}\right) \mathcal{O}\left(\eta^{2j}\right) + \mathrm{h.c.},
\end{split}
\end{equation}
with $j\geq 1$. Hence, depending on $\Delta\phi$, all even ($\Delta\phi=0$) or all odd ($\Delta\phi=\pi$) terms in $\eta$ are suppressed. In the main text, we neglect all higher-order terms in $\eta$ and obtain Eq.~\eqref{eq:monochrmatic_sw} in the main text.
\section{Phase stabilization scheme}
\begin{figure}[ht!]
    \centering
    \includegraphics[width=\linewidth]{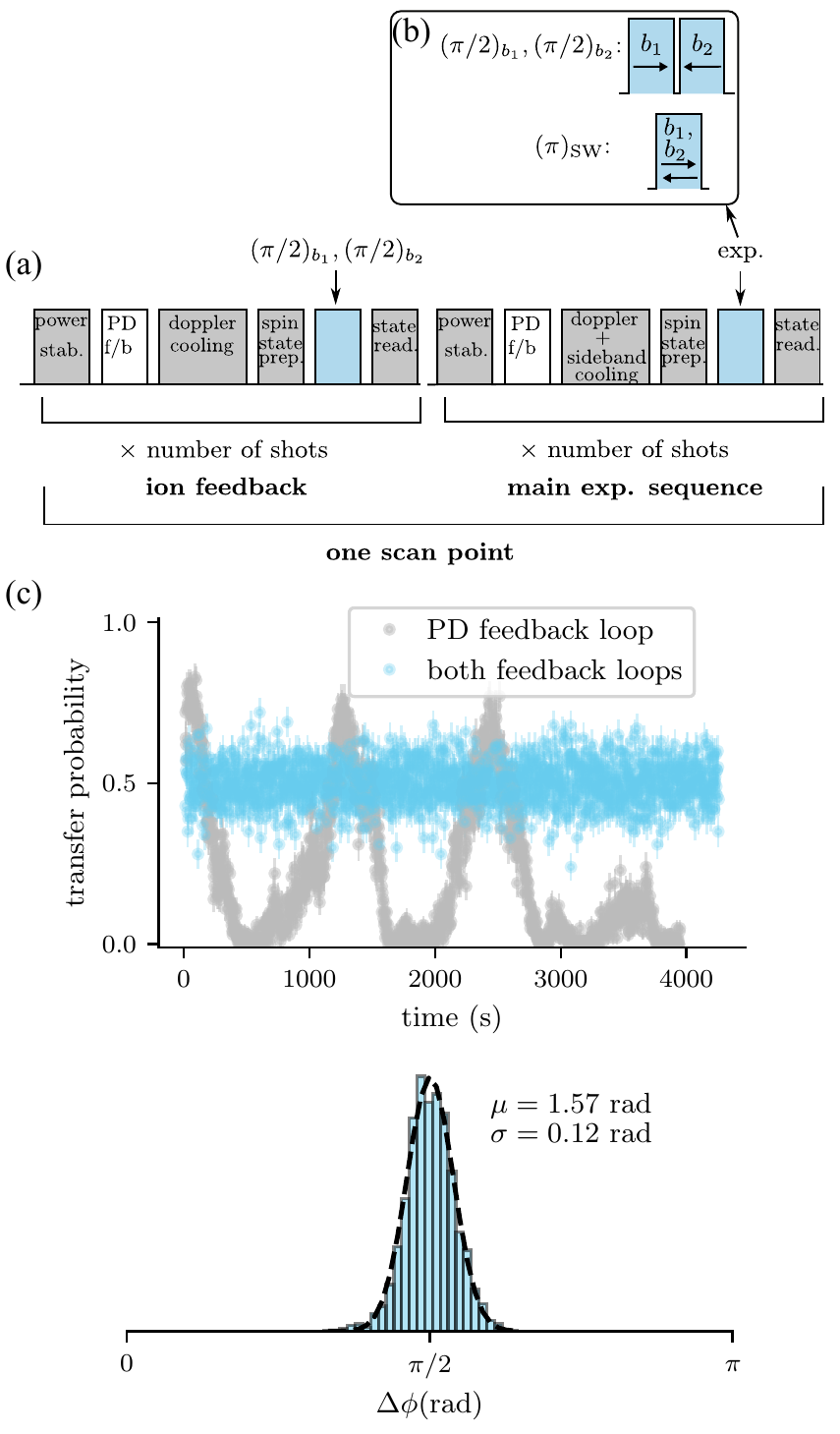}
    \caption{(a)
    Every scan point consists of $N=100$ shots of the main experimental sequence preceded by $M=100$ shots of the ion feedback. Both the main experimental sequence shot and the ion feedback shot start with stabilizing the interferometer with respect to the photodiode lock point.
    (b) Inset showing the pulse sequence for typical main experiments (e.g.~the pulse sequence for the zero-delay Ramsey experiment and the monochromatic SW pulse). 
    (c) Phase stability data measured with the ion when only the PD feedback is enabled vs. when both feedback loops are enabled. The transfer probability values are converted to phase fluctuations and shown in the histogram. We use a Gaussian fit to determine the rms deviation, $\sigma_\phi=\sigma$.  }
    \label{fig:feedback_pulse_sequence}
\end{figure}
We actively stabilize the phase of the SW with respect to the position of the ion(s). Two feedback loops are employed for this.
Firstly, we eliminate fast drifts. 
By picking off a small fraction of light from each of the beams $b_1$, $b_2$, and temporarily shifting the frequency of $b_2$, we create a heterodyne signal of the intensity interference. This is measured with a photodiode (PD) [Fig.~\ref{fig:setup}(a)] and used to infer the optical phase correction. We apply this feedback loop for $\SI{50}{\micro\second}$ before each shot of the experiment, indicated as ``PD f/b" in Fig.~\ref{fig:feedback_pulse_sequence}. Secondly, because the PD lock point is around $\approx30$~cm away from the location of the ions, we add a second feedback loop using the ion itself as a sensor. We do this by performing a $\pi/2$-pulse using $b_1$ ($\pi/2|_{b_1}$) followed immediately by a $\pi/2$-pulse using $b_2$ ($\pi/2|_{b_2}$). This is equivalent to a zero-delay Ramsey sequence, which gives a signal sensitive to the difference in phase between the two pulses, hence the relative phase between the beams. For all the SW experimental data shown in the main text, we interleave the $\pi/2|_{b_1}, \pi/2|_{b_2}$ feedback sequence with the main experiment for each scan point. Each is repeated for 100 shots. Hence, the ion feedback sequence is repeated every $\SI{0.5}{\second}$. In Fig.~\ref{fig:feedback_pulse_sequence}(c), we show the signal measured on the ion using the $\pi/2|_{b_1}, \pi/2|_{b_2}$ sequence when only the PD feedback is enabled (grey) and when both feedback loops are enabled (cyan). We infer residual phase fluctuations by converting the transfer probability data to the phase difference $\Delta\phi$ and determine the rms deviation, $\sigma_{\Delta\phi} = \SI{0.12}{\radian}$. We express this in terms of the standing wave period $\sigma_{\Delta\phi}/(2\pi) = \lambda_{\rm SW}/50$ or the wavelength of 674-nm light $\lambda/100$.

\section{Calibration of the anti-node position of bichromatic fields}
\begin{figure}[ht]
    \centering
    \includegraphics[]{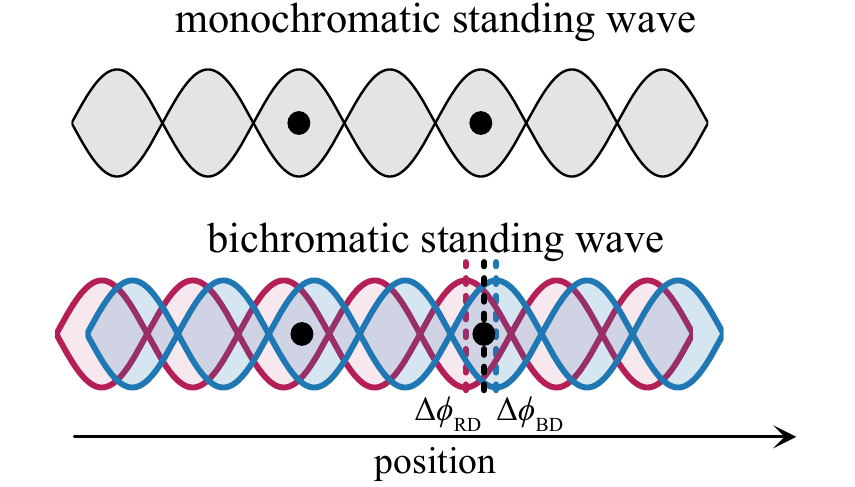}
    \caption{Illustration of the phase offset occurring when applying a bichromatic off-resonant standing wave (lower row) relative to the phase-stabilized monochromatic field at qubit frequency (upper row). The SWs forming the bichromatic field are offset by $\Delta \phi_{\mathrm{RD}}$ and $\Delta \phi_{\mathrm{BD}}$ relative to the ion position respectively.}
    \label{fig:bichromatic_lattice}
\end{figure}

Experimentally we found that there is a significant frequency dependence on the phase acquired in the electronic signal chain. This results in the two SWs that form the bichromatic field being offset with respect to each other and with respect to the ions (Fig.~\ref{fig:bichromatic_lattice}). We calibrate this offset for each SW separately by applying a monochromatic SW pulse on two ions while scanning $\Delta\phi$ and being off-resonant from the qubit resonance by $\delta =\pm( \omega_z - \delta_g)$ as required by the MS interaction. From the resulting dynamics, which are similar to those shown in Fig.~\ref{fig:lattice-two-ion-rotations}, the shift of the SW anti-node can be extracted. We can calibrate this with an accuracy of $\Delta\phi_{\rm bi} = \SI{4.2e-2}{\radian}$. 

\section{Matching the ion spacing to the standing wave periodicity}
\begin{figure}[ht]
    \centering
    \includegraphics[]{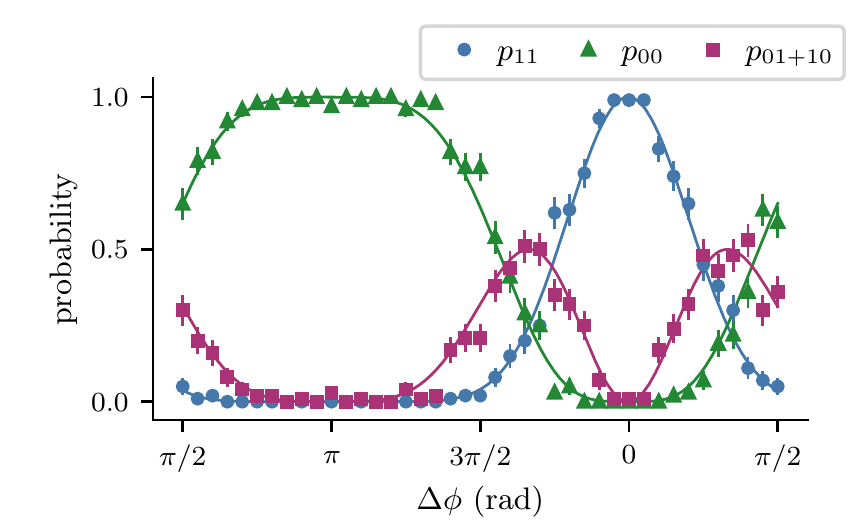}
    \caption{Monochromatic SW interacting with two ions. Final state probability of two-ion bright $p_{11}$, one-ion bright $p_{01}+p_{10}$ and no-ion bright $p_{00}$ as a function of the SW phase at the position of the ions, while the SW is on resonance with the carrier when the ion internal states are initialized such that $p_{11}=1$. The SW pulse duration $t_p$ is set such that complete population transfer is achieved at maximal carrier coupling.}
    \label{fig:lattice-two-ion-rotations}
\end{figure}

In our apparatus we address the ions with a global SW angled at \SI{45}{\degree} relative to the linear crystal axis [Fig.~\ref{fig:setup}(a)]. To ensure that the ions have the same coupling ratio between carrier and sideband we must position them such that they see the same SW phase [Fig.~\ref{fig:bichromatic_lattice} (upper row)].
Hence, we need to adjust their spacing such that the distance between the ions, projected onto the $\bm{k}$-vector of the SW, is an integer multiple of the SW periodicity.
We perform the same experiment as in Fig.~\ref{fig:lattice-single-rotations}(a) but on two ions (Fig.~\ref{fig:lattice-two-ion-rotations}). 
We extract the relative difference between SW periodicity and ion spacing for a given axial confinement by fitting the resulting dynamics (two-ion bright, single-ion bright, and no-ion bright population). Based on this, we match the spacing by adjusting the axial confinement strength. We can calibrate the phase-match between the two ions with an accuracy of $\Delta\phi_{\rm sp} = \SI{3.3e-2}{\radian}$. Finally, the two ions are spaced by $\approx\SI{3.8}{\micro \meter}\cdot\cos(45^\circ)=4\lambda$.

N.b.~addressing a linear ion crystal consisting of more than two ions requires more consideration. In a harmonic potential, the ions will not be equally spaced for more than three ions. To be able to place all ions at the same SW phase and individually address an arbitrary number of them, a possible architecture would be an array of tightly focused standing waves whose individual $\bm{k}$-vectors are along the radial direction of the linear crystal. Hence, a separate SW is created for every addressed ion.

\section{Standing wave fidelity error estimates}
In this section, we present analytic estimates for the errors on the two-qubit gates due to the SW. We consider the errors due to amplitude imbalance between the two beams $b_1$ and $b_2$ (visibility error); the quality of the SW phase stabilization which introduces a position jitter of the SW relative to the ions (phase stability error); mismatched spacing of the ions relative to the SW periodicity $\lambda_\mathrm{SW}$ (ion spacing error); and phase misalignment of the blue and red detuned SWs, i.e. $\Delta\phi_\mathrm{BD}\neq 0\ {\rm or}\ \Delta\phi_\mathrm{RD}\neq0$ (phase misalignment in the bichromatic SW error). The error contributions are summarized in Table~\ref{tab:SW_errors}.
\begin{table}[ht]
\caption{\label{tab:SW_errors}
Estimated errors for the SW-MS gate for the fastest gate duration (\SI{15}{\micro\second}) attempted. Errors common to both the TW and SW are excluded.}
\begin{ruledtabular}
\begin{tabular}{lcSc}
{Error source}& {Fluctuation~in}  & \multicolumn{2}{c}{$\mathrm{Error}/ \num{e-4}$}\\
{}&{parameter}&{square}&{shaped}\\
\colrule
Visibility carrier& $\Delta\Omega/\Omega = \num{0.05}$ &  3.46 & 0.00 \\
Phase carrier& $\sigma_{\Delta\phi} = \SI{0.12}{\radian}$ & 61.0 & 0.06\\
Phase sideband& $\sigma_{\Delta\phi}= \SI{0.12}{\radian}$ &0.03 & 0.03\\
Ion spacing carrier & $\Delta\phi_\mathrm{sp} = \SI{0.033}{\radian}$& 2.12& 0.00 \\
RD/BD phase mismatch & $\Delta\phi_\mathrm{bi} = \SI{0.042}{\radian}$&  15.4& 0.02\\
\colrule
Total error&& 82.0 & 0.11
\end{tabular}
\end{ruledtabular}
\end{table}

We estimate the error of an unwanted unitary $\hat{H}_\textrm{err}$ by considering the effect of its propagator $\hat{U}_\textrm{err}$ on the initial state $\ket{u,0}$.
For a single ion:
\begin{align}\label{eq:single_ion_err}
    \epsilon &= 1- |\bra{u,0}\hat{U}_\textrm{err}\ket{u,0}|^2 = 1- \cos^2\theta =\frac{1}{2}(1-\cos2\theta), 
\end{align}
where
\begin{equation}\label{eq:theta}
\theta = \left|\frac{1}{\hbar} \int \hat{H}_\textrm{err}(t)dt \right|.
\end{equation}
For two ions we neglect cross-contributions and obtain
\begin{equation}\label{eq:two_qubit_err}
    \epsilon \leq 1-\cos^2\theta_1 \cos^2\theta_2 \stackrel{\theta_1=\theta_2}{=} 1 -\cos^4\theta \stackrel{\theta \ll \pi/2}{=} 2\theta^2.
\end{equation}

\subsection{Visibility error}
We assume that the two counter-propagating bichromatic fields, used for implementing the SW-MS, have Rabi frequencies $\Omega \pm \Delta\Omega/2$.
\begin{align}
	\hat{H}_{\rm SW-MS} &= \hbar \left(\Omega +\frac{\Delta\Omega}{2} \right)e^{i\eta(\hat{a}+\hat{a}^\dagger)} \hat{S}_+ \cos(\delta t)e^{i\phi_1} \nonumber \\
	&-\hbar \left(\Omega -\frac{\Delta\Omega}{2}\right)e^{-i\eta(\hat{a}+\hat{a}^\dagger)} \hat{S}_+ \cos(\delta t)e^{i\phi_2} +\mathrm{h.c.}   \\
	&= 2\hbar \Omega e^{i\tilde{\phi}} \sin(\eta(\hat{a}+\hat{a}^\dagger) +\Delta\phi)\cos(\delta t)\hat{S}_+ \nonumber  \\
		&+\hbar \Delta\Omega e^{i\tilde{\phi}-\frac{\pi}{2}} \cos(\eta(\hat{a}+\hat{a}^\dagger) +\Delta\phi)\cos(\delta t) \hat{S}_+ +\mathrm{h.c.}
\end{align}
Hence, the error is caused by:
\begin{equation}
	\hat{H}_\textrm{err} =\hbar \Delta\Omega \cos(\delta t) \hat{S}_{\tilde{\phi}-\frac{\pi}{2}}
\end{equation}
and consequently, following Eq.~\eqref{eq:theta} and \eqref{eq:two_qubit_err}:
\begin{align}
    \theta & = \int_{0}^{t_f} \Delta\Omega \cos(\delta t) dt = \frac{\Delta \Omega}{\delta} \sin(\delta t)|_0^{t_f} \leq \frac{\Delta\Omega}{\delta}\\
    \epsilon & \leq 2 \left(\frac{\Delta\Omega}{\delta}\right)^2. 
\end{align}
\subsection{Phase stability error}
For the phase stability, we assume $\Delta\phi$ is a random variable sampled from a Gaussian distribution with mean $0$ and variance $\mathrm{Var}(\Delta\phi) = \sigma_{\Delta\phi}^2$.
The mean is 0 as the ions are placed at the anti-nodes for the gates.
Using the SW-MS Hamiltonian in Eq.~\eqref{eq:H_MS_sw}, we emphasize two errors. We consider small variations in $\Delta\phi$.
The second term gives rise to an off-resonant carrier coupling:
\begin{equation} \label{eq:carrier_phase_err}
\hat{H}_\textrm{err} =2\hbar \Omega\hat{S}_{\tilde{\phi}} \cos{(\delta t)}\left( \frac{\Delta\phi}{2} \right). 
\end{equation}

Using Eq.~\eqref{eq:theta} and~\eqref{eq:two_qubit_err} and averaging over the Gaussian fluctuations of $\Delta\phi$, we infer:
\begin{align}
    \mathrm{Var}(\theta) &\leq \left(\frac{2\Omega}{\delta}\right)^2\ \mathrm{Var}\left(\frac{\Delta\phi}{2}\right) \\
    \epsilon &\leq 2 \left(\frac{2\Omega}{\delta}\right)^2\ \mathrm{Var}\left(\frac{\Delta\phi}{2}\right).
\end{align}
The first term in Eq.~\eqref{eq:H_MS_sw} gives a modulation on the sideband coupling:
\begin{align}
    \hat{H}_\textrm{err} &= 2\hbar\eta\Omega \hat{S}_{\tilde{\phi}} \cos{(\delta t)}(\hat{a}e^{-i \omega_z t} + \hat{a}^\dagger e^{i \omega_z t})\left(\frac{\Delta\phi}{2} \right)^2\label{eq:sb_err}\\
    \epsilon &\leq 3\left(\frac{\Omega\eta}{\delta_g}\right)^2\ \mathrm{Var}\left(\frac{\Delta\phi}{2}\right)^2.
\end{align}

In the derivations above, we assumed $\mathrm{Var}(\Delta\phi)\ll 1$ and used that for a Gaussian distribution the expected value of Eq.~\eqref{eq:two_qubit_err} becomes:
\begin{align}
    \epsilon \leq 2\mathrm{Var}(\theta). \nonumber
\end{align}
\subsection{Ion spacing error}
We assume that there is a $\Delta\phi_{\rm sp}$ mismatch between the SW periodicity and the ion spacing. This means that we will not be able to position the ions such that they both experience $\Delta\phi = 0$. Here, we will only consider the error due to the off-resonant carrier term; the spin-motional coupling is more robust as a result of the quadratic dependence on $\Delta\phi$ [Eq.~\eqref{eq:sb_err}]. The largest error due to the carrier occurs when $\Delta\phi = 0$ at one ion and $\Delta\phi = \Delta\phi_{\rm sp}$ at the other ion.
Using the single-ion error in Eq.~\eqref{eq:single_ion_err} and Eq.~\eqref{eq:carrier_phase_err}, we infer:
\begin{align}
   \epsilon \leq \left(\frac{2\Omega}{\delta}\right)^2\  \left(\frac{\Delta\phi_{\rm sp}}{2}\right)^2.
\end{align}

\subsection{Phase misalignment in the bichromatic SW
error}
We re-write Eq.~\eqref{eq:H_MS_sw} for $\Delta\phi_{\rm BD}\neq \Delta\phi_{\rm RD}$: 
\begin{equation}
    \begin{split}
    \hat{H}_\mathrm{SW-MS} =\ &\hbar\Omega\eta \hat{S}_{+}e^{-i\delta t}e^{i\tilde{\phi}}(\hat{a}e^{-i \omega_z t} + \hat{a}^\dagger e^{i \omega_z t})\cos{\left(\frac{\Delta\phi_{\rm BD}}{2} \right)}\\    
    +&\hbar \Omega\hat{S}_{+} e^{-i\delta t} e^{i\tilde{\phi}} \sin{ \left(\frac{\Delta\phi_{\rm BD}}{2} \right)}\\
    +&\hbar\Omega\eta \hat{S}_{+} e^{i\delta t}e^{i\tilde{\phi}}(\hat{a}e^{-i \omega_z t} + \hat{a}^\dagger e^{i \omega_z t})\cos{\left(\frac{\Delta\phi_{\rm RD}}{2} \right)}\\    
    +&\hbar \Omega\hat{S}_{+}  e^{i\delta t}e^{i\tilde{\phi}} \sin{ \left(\frac{\Delta\phi_{\rm RD}}{2} \right) + \textrm{h.c.},}
    \end{split}
\end{equation}
where we assumed used that the tones are detuned by $\pm \delta$ respectively and $\delta\approx\omega_z$.
Based on the considerations mentioned above, we only retain the error due to the carrier term:
\begin{equation}
    \begin{split}
    \hat{H}_\mathrm{err} =\ &\hbar \Omega\hat{S}_{+}  e^{-i\delta t} e^{i\tilde{\phi}}\left(\frac{\Delta\phi_{\rm BD}}{2} \right)
    +\hbar \Omega\hat{S}_{+}  e^{i\delta t}e^{i\tilde{\phi}} \left(\frac{\Delta\phi_{\rm RD}}{2} \right)\\
    +&\hbar \Omega\hat{S}_{-}  e^{i\delta t} e^{-i\tilde{\phi}}\left(\frac{\Delta\phi_{\rm BD}}{2} \right)
    +\hbar \Omega\hat{S}_{-}  e^{-i\delta t}e^{-i\tilde{\phi}} \left(\frac{\Delta\phi_{\rm RD}}{2} \right),
    \end{split}
\end{equation}
\begin{equation}
    \epsilon \leq 2\left(\frac{2\Omega}{\delta} \right)^2\left(\left(\frac{\Delta\phi_{\rm BD}}{2} \right)^2 +  \left(\frac{\Delta\phi_{\rm RD}}{2} \right)^2\right).
\end{equation}
We assume $\Delta\phi_{\rm RD} = \Delta\phi_{\rm BD} = \Delta\phi_{\rm bi}$.
Then,
\begin{equation}
    \epsilon \leq 4\left(\frac{2\Omega}{\delta} \right)^2\left(\frac{\Delta\phi_{\rm bi}}{2} \right)^2.
\end{equation}

\subsection{Pulse shaping}
In Table~\ref{tab:SW_errors} the errors for the square pulse are inferred based on the derivations above. Smoothly ramping the amplitude of the electric field (referred to as pulse shaping in the main text) over a duration of $t_R =\SI{10}{\micro\second}$ suppresses the contribution of error terms originating from off-resonant carrier coupling. The phase acquired by an error is then modified
\begin{equation}
\theta_{\textrm{shaped}} = \left|\frac{1}{\hbar} \int_0^{t_f} g(t)\hat{H}_\textrm{err}(t)dt \right|,
\end{equation}
where $g(t)$ is the pulse shape
\begin{equation}
g(t) = 
\begin{cases}
\sin(\pi t/2t_{ R})^2,\ t<t_R\\
1,\ t_R\leq t \leq t_f-t_R\\
\sin(\pi (t_f -t)/2t_{ R})^2,\ t_f-t_R< t<t_f.
\end{cases}
\end{equation}

We then define the error ratio between the shaped and the square pulse as
\begin{equation}
    r = \left| \theta_{\textrm{shaped}} \right|^2/ \left| \theta_{\textrm{square}}\right|^2,
\end{equation}
where we have used Eq.~\eqref{eq:two_qubit_err}.
We find that $r \leq \num{1e-3}$ for our chosen experimental parameters. We have applied this suppression factor to the carrier-related error estimates in the pulse shaped column of Table~\ref{tab:SW_errors}.
\label{suppl_material}

\end{document}